\documentclass[aps,prd,superscriptaddress,nofootinbib,onecolumn,10pt,notitlepage,preprintnumbers]{revtex4-1}

\usepackage{amsmath,amssymb,amsbsy,amstext,amsfonts,mathrsfs,latexsym,bm}

\usepackage[colorlinks=true,urlcolor=blue,anchorcolor=blue,citecolor=blue,linkcolor=blue,linktocpage=true,pdfa=true]{hyperref}

\newcommand{\mpl}{M_{\scriptscriptstyle\mathrm{Pl}}}
\newcommand{\mplt}{\tilde M_{\scriptscriptstyle\mathrm{Pl}}}

\begin{document}

\title{Cuscuton Gravity as a Classically Stable Limiting Curvature Theory}

\author{Jerome Quintin}
\email{jerome.quintin@aei.mpg.de}
\affiliation{Max Planck Institute for Gravitational Physics (Albert Einstein Institute), D-14476 Potsdam, Germany}

\author{Daisuke Yoshida}
\email{dyoshida@hawk.kobe-u.ac.jp}
\affiliation{Department of Physics, Kobe University, Kobe 657-8501, Japan}

\preprint{KOBE-COSMO-19-19}

\begin{abstract}
 Finding effective theories of modified gravity that can resolve cosmological singularities and avoid other physical pathologies such as ghost and gradient instabilities has turned out to be a rather difficult task.
 The concept of limiting curvature, where one bounds a finite number of curvature-invariant functions thanks to constraint equations, is a promising avenue in that direction, but its implementation has only led to mixed results.
 Cuscuton gravity, which can be defined as a special subclass of $k$-essence theory for instance, is a minimal modification of gravity since it does not introduce any new degrees of freedom on a cosmological background.
 Importantly, it naturally incorporates the idea of limiting curvature.
 Accordingly, models of cuscuton gravity are shown to possess non-singular cosmological solutions and those appear stable at first sight.
 Yet, various subtleties arise in the perturbations such as apparent divergences, e.g., when the Hubble parameter crosses zero.
 We revisit the cosmological perturbations in various gauges and demonstrate that the stability results are robust even at those crossing points, although certain gauges are better suited to analyze the perturbations.
 In particular, the spatially-flat gauge is found to be ill defined when $H=0$.
 Otherwise, the sound speed is confirmed to be generally close to unity in the ultraviolet, and curvature perturbations are shown to remain essentially constant in the infrared throughout a bounce phase.
 Perturbations for a model of extended cuscuton (as a subclass of Horndeski theory) are also studied and similar conclusions are recovered.
\end{abstract}

\maketitle

\section{Introduction}

One of the outstanding implications of General Relativity (GR) is the inevitable presence of singularities in spacetime.
This is demonstrated by the singularity theorems of Penrose and Hawking \cite{PenroseHawking} and represents a fundamental limitation of GR.
A theory of gravity going beyond GR should thus be able to yield a better description of the physics at extremely high curvature scales and ultimately resolve singularities altogether.

However, building modified gravity theories that avoid the classical singularities of GR in a theoretically-consistent manner has been shown to be a rather difficult task.
In an attempt at modifying gravity to effectively describe gravitation up to high curvature scales and resolve spacetime singularities, higher-curvature terms can be added to the Einstein-Hilbert action.
Then, the hope is that the resulting theory serves as an effective theory of the underlying unified theory of quantum gravity.
For example, $\alpha'$ corrections in low-energy effective string theory add higher-curvature terms to the action \cite{ST} and are believed to represent the low-energy limit of full string theory.
However, higher-curvature terms do not guarantee the absence of spacetime singularities.
A better approach is thus to ensure that the higher-curvature terms are added to the action in such a way that those curvature terms are bounded.
Furthermore, if the theory can be engineered such that known non-singular spacetimes are obtained asymptotically, then any curvature-invariant function can be bounded, ensuring the full spacetime is non-singular.
This is the principle of limiting curvature\footnote{Limiting curvature can also be implemented to avoid future singularities (see, e.g., \cite{futureLC}).} \cite{earlyLC,Mukhanov:1991zn,Brandenberger:1993ef}.

Early constructions applying the limiting curvature principle used Lagrange multipliers in the action such that specific constraint equations followed.
The role of these constraint equations is precisely to bound curvature-invariant functions.
For example, a term of the form $\varphi I-V(\varphi)$ in the Lagrangian density yields a constraint equation of the form $I=\partial V/\partial\varphi$ upon variation with respect to the Lagrange multiplier $\varphi$.
Here, $I$ is meant to be a function of curvature invariants built out of the Riemann tensor, contractions, and derivatives thereof, such as $R$, $R_{\mu\nu}R^{\mu\nu}$, $\nabla_\mu R_{\alpha\beta}\nabla^\mu R^{\alpha\beta}$, etc.
Then, if the potential $V(\varphi)$ is chosen appropriately, i.e., such that $\partial V/\partial\varphi$ remains finite for all physically admissible values of $\varphi$, the curvature-invariant function $I$ is guaranteed to be bounded.
Limiting a finite number of curvature invariants, it is then possible to avoid singularities altogether by choosing the appropriate `boundary conditions', meaning having a spacetime manifold that is asymptotically non-singular.
In cosmology, such constructions could start and end in a non-singular de Sitter spacetime, while being Friedmann-Lema\^itre-Robertson-Walker (FLRW) cosmology in between \cite{Brandenberger:1993ef,Yoshida:2017swb,Yoshida:2018ndv}.
This is a surprising result because cosmologies starting with a phase of accelerated expansion are usually past geodesically incomplete \cite{inflationSingular,Yoshida:2018ndv,Numasawa:2019juw}.
Yet, with limiting curvature, one can make sure that the spacetime is truly extendible beyond its past boundary and exempt of all types of singularities \cite{Yoshida:2018ndv}.

While the prospect of avoiding singularities with limiting curvature is good, the models of limiting curvature that use higher-curvature terms in the action come with a severe issue.
Indeed, they are usually plagued with instabilities at the level of the cosmological perturbations \cite{Yoshida:2017swb}, or when dealing with black hole spacetimes, divergences can remain even at the background level \cite{Yoshida:2018kwy}.

There exist other implementations of the limiting curvature principle that do not rely on higher-curvature terms in the action.
For example, mimetic gravity \cite{mimeticgeneral} has been shown to be able to implement limiting curvature, thus allowing for non-singular spacetime constructions, such as non-singular bouncing cosmology \cite{mimeticbounce} (for a different implementation, see \cite{mimeticbounce2}), non-singular asymptotically free cosmology \cite{Chamseddine:2019bcn} or non-singular black hole spacetime \cite{mimeticBH}.
In the context of mimetic gravity, one still has a Lagrange multiplier yielding a constraint equation, and if appropriately set up, the constraint equation can ensure the finiteness of a given geometrical quantity.
Rather than using higher-curvature terms, mimetic gravity uses a scalar field to modify GR.
Yet, mimetic gravity cannot escape instabilities in cosmological perturbations \cite{mimeticinst}, in particular in a non-singular cosmological bounce \cite{Ijjas:2016pad}.

In this paper, we claim that there exists another form of modified gravity that can implement limiting curvature, but that can safely avoid instabilities in a non-singular cosmological setup.
The concept of cuscuton gravity \cite{Afshordi:2006ad,Afshordi:2007yx,Afshordi:2009tt,Afshordi:2010eq,Bhattacharyya:2016mah,Iyonaga:2018vnu} arises when trying to minimally modify GR by introducing a scalar field that does not propagate any new degrees of freedom.
Indeed, one can setup a scalar-tensor theory of gravity where the scalar field satisfies at most a first-order background equation of motion (the constraint equation) and similarly where no kinetic term appears in the scalar perturbed action (in vacuum).
The most basic cuscuton theory has a Lagrangian density of the form $\sqrt{-\partial_\mu\phi\partial^\mu\phi}-V(\phi)$ and has been shown to possess many interesting properties.
Indeed, the theory has stable cosmological perturbations \cite{Afshordi:2007yx,Boruah:2017tvg}, is free from caustic instabilities \cite{deRham:2016ged}, appears as the extreme relativistic limit of a 5-dimensional brane theory \cite{Chagoya:2016inc} and as the ultraviolet (UV) limit of an anti-Dirac-Born-Infeld theory \cite{Afshordi:2016guo}, and has non-singular bouncing solutions \cite{Boruah:2018pvq} (see also Ref.~\cite{Romano:2016jlz}), besides other phenomenological applications \cite{cuscutonpheno}.
Also, a Hamiltonian analysis showed that it propagates only two degrees of freedom in the unitary gauge, hence to all orders in perturbation theory no new degrees of freedom appear beyond the usual two polarization states of gravitational waves \cite{Gomes:2017tzd}.
In fact, it is probably true in any gauge since the extra mode becoming apparent outside the unitary gauge\footnote{Even if one works in a gauge where $\partial_i\phi\neq 0$, one assumes that $\partial_\mu\phi$ remains a timelike vector in cuscuton gravity.
Non-dynamical scalar fields with spacelike rather than timelike gradient vectors can also lead to interesting scalar-tensor theories of modified gravity \cite{Gao:2018izs}.} appears to be a `shadowy' mode, which does not propagate \cite{DeFelice:2018mkq}.
Cuscuton gravity is a subclass of such theories having only two degrees of freedom, yet being apparently different from GR \cite{mingrav,Mukohyama:2019unx}.
More formally, it has also been shown that such a theory has infinitely many new symmetries \cite{Pajer}.

In this paper, our goal is to first show how cuscuton gravity can naturally implement the limiting curvature principle and yield a non-singular bouncing cosmology.
Then, the main novelty consists in revisiting cosmological perturbations from the perspective of many different gauges.
Indeed, while stability through a non-singular bounce has been claimed in the unitary gauge \cite{Boruah:2018pvq}, it is often the case that there remain points in time where divergences can occur.
For instance in non-singular bounces constructed from Horndeski theory \cite{Horndeskibounce}, there exists a point in time where a certain quantity\footnote{In a $k$-essence theory or generally without kinetic braiding \cite{Deffayet:2010qz}, this quantity is the Hubble parameter. Thus, the issue arises when $H=0$, which corresponds to the bouncing point when the universe transitions from contraction to expansion.} appearing in many denominators crosses zero \cite{Battarra:2014tga,Quintin:2015rta}.
The issue is resolved by looking at different gauges and reduces to a technical subtlety when approached with care \cite{gammacrossing}.
Nevertheless, it is critical when addressing the question of stability in a non-singular spacetime.
Indeed, there has been claims of no-go theorems proving the impossibility of obtaining stable non-singular cosmologies in Horndeski theory \cite{Libanov:2016kfc,Kobayashi:2016xpl,Akama:2017jsa,EFTnogo}.
Whether certain models can evade the no-go theorems has to be properly shown.
In this work on cuscuton gravity, we demonstrate that there exist special points in time that can appear problematic, just as in Horndeski theory\footnote{In fact, cuscuton gravity is often implemented within Horndeski theory, i.e., it is a special subclass. The simplest models of cuscuton gravity are of the $k$-essence type.}.
We show, however, that it is just a question of choosing the appropriate gauge in certain conditions.
Accordingly, stability of a limiting curvature non-singular bounce is confirmed.
We show this result not only in simple cuscuton gravity (in the subclass of $k$-essence theory), but also for the recently proposed extended cuscuton \cite{Iyonaga:2018vnu}, i.e., when it is a subclass of Horndeski theory.
Besides stability, we comment on how perturbations generally behave passing through a non-singular bounce in cuscuton gravity.
In particular, we find that large-wavelength curvature perturbations cannot grow more than by a certain amount (as guaranteed by the stability), such that the constant mode remains the dominant one throughout the bounce phase.
We discuss the implications of this result.

The paper is organized as follows.
We start in Sec.~\ref{sec:cuscutonreview} by reviewing cuscuton gravity, in its original form as a subclass of $k$-essence theory and as a minimally-modified theory of gravity.
In particular, we show how the theory implements the concept of limiting curvature and how it can yield a non-singular bouncing cosmology.
In Sec.~\ref{sec:cosmopert}, we revisit the cosmological perturbations in this context, demonstrating that the theory is free of ghost and gradient instabilities throughout cosmic time.
We explore various gauges, indicating when they are well defined and when they are not.
We then comment on the behavior of the whole spectrum of perturbations in the bounce phase.
The range of the sound speed is analyzed, and an upper bound on the growth of infrared curvature perturbations is found.
Section \ref{sec:extendedcuscuton} broadens the analysis to a model of extended cuscuton.
As before, the stability and finiteness of the perturbations is examined.
The results are summarized and further discussed in Sec.~\ref{sec:fin}.
We use the metric signature $(-,+,+,+)$, and the reduced Planck mass is defined to be $\mpl\equiv 1/\sqrt{8\pi G_\mathrm{N}}$, where $G_\mathrm{N}$ is Newton's gravitational constant.

\section{Review of Cuscuton gravity and non-singular bouncing cosmology}\label{sec:cuscutonreview}

\subsection{Cuscuton as a special subclass of k-essence}

Let us introduce the cuscuton. The cuscuton is defined as a scalar field $\phi$ that is non-dynamical at the background level and that has no propagating scalar linear perturbation. In other words, the cuscuton field must have at most a first-order equation of motion in the unitary gauge (i.e., when $\phi(t,\mathbf{x})=\phi(t)$, so $\delta\phi(t,\mathbf{x})=0$) and a vanishing kinetic term in its second-order perturbed action.
There are many Lagrangians that can satisfy these requirements, but let us start by introducing the simplest possibility. Let us start from a generic $k$-essence scalar field \cite{ArmendarizPicon:2000ah} with action
\begin{equation}
 S=\int\mathrm{d}^4x\,\sqrt{-g}\left(\frac{\mpl^2}{2}R+P(X,\phi)\right)\,,
\end{equation}
where $X\equiv -g^{\mu\nu}\nabla_\mu\phi\nabla_\nu\phi/2$.
Upon variation with respect to $\phi$, the generic equation of motion is
\begin{equation}
 g^{\mu\nu}\nabla_\mu(P_{,X}\nabla_\nu\phi)+P_{,\phi}=0\,,
\end{equation}
where a coma denotes a partial derivative, e.g., $P_{,X}\equiv\partial P/\partial X$.
Let us now consider an FLRW background metric of the form
\begin{equation}
 g_{\mu\nu}\mathrm{d}x^\mu\mathrm{d}x^\nu=-\mathrm{d}t^2+a(t)^2\delta_{ij}\mathrm{d}x^i\mathrm{d}x^j\,.
\end{equation}
Then, in the unitary gauge where we have $X=\dot\phi^2/2$, the equation of motion reduces to
\begin{equation}
 (P_{,X}+2XP_{,XX})\ddot\phi+3HP_{,X}\dot\phi+P_{,X\phi}\dot\phi^2-P_{,\phi}=0\,,
\end{equation}
where $H\equiv\dot a/a$ is the Hubble parameter.
For $\phi$ to be a cuscuton field, one must impose that
\begin{equation}
 P_{,X}+2XP_{,XX}=0\iff P(X,\phi)=C_1(\phi)\sqrt{|X|}+C_2(\phi)\,,
\end{equation}
where $C_1$ and $C_2$ are arbitrary functions of $\phi$.
We note that the solution on the right-hand side is valid for all values of $X$.
A common convention is to write the cuscuton Lagrangian as
\begin{equation}
 \mathcal{L}=P(X,\phi)=\pm M_L^2\sqrt{2X}-V(\phi)\,,
 \label{eq:Lcusc}
\end{equation}
which can be done by performing a field redefinition of the form $\phi\rightarrow f(\phi)$ such that $C_1(f(\phi))f_{,\phi}=\pm\sqrt{2}M_L^2$ and by writing $C_2(f(\phi))\equiv -V(\phi)$.
Also, we assume that $X>0$, i.e., $\partial_\mu\phi$ is a timelike vector, which is always the case in the unitary gauge.
Note that $M_L$ is a dimensionful parameter with mass dimension.
With such a Lagrangian, it becomes clear that the background equation of motion is non-dynamical and reduces to a constraint equation:
\begin{equation}
 \mp\mathrm{sgn}(\dot\phi)3M_L^2H=V_{,\phi}\,.
 \label{eq:constreqFLRW}
\end{equation}
This constraint equation is of the form of a limiting curvature constraint since choosing an appropriate potential $V(\phi)$ such that $V_{,\phi}$ is finite for all values of $\phi$ ensures $H$ is bounded and does not diverge. In particular, the mass scale $M_L$ is associated with the limiting curvature scale.

This is still true in a more general background (i.e., not necessarily FLRW). The Euler-Lagrange equation for the Lagrangian \eqref{eq:Lcusc} is
\begin{equation}
 \pm M_L^2g^{\mu\nu}\nabla_\mu\left(\frac{\nabla_\nu\phi}{\sqrt{2X}}\right)=V_{,\phi}\,.
 \label{eq:constreqgen}
\end{equation}
One can view the cuscuton field as a perfect fluid with energy-momentum tensor
\begin{equation}
 T_{\mu\nu}=(\rho+p)u_\mu u_\nu+pg_{\mu\nu}=Pg_{\mu\nu}+P_{,X}\nabla_\mu\phi\nabla_\nu\phi=\left(\pm M_L^2\sqrt{2X}-V(\phi)\right)g_{\mu\nu}\pm M_L^2\frac{\nabla_\mu\phi\nabla_\nu\phi}{\sqrt{2X}}\,,
\end{equation}
where
\begin{equation}
 \rho=2XP_{,X}-P\,,\qquad p=P\,,\qquad u_\mu=\pm\frac{\nabla_\mu\phi}{\sqrt{2X}}\,,
\end{equation}
are the energy density, pressure, and fluid velocity, respectively.
Geometrically, $u_\mu$ is also the normal, unit vector to a constant-$\phi$ hypersurface.
Such a surface has an extrinsic curvature tensor $K_{\mu\nu}$ with trace (the mean scalar curvature) given by $K=\nabla_\mu u^\mu$. This is precisely the quantity that enters the constraint equation \eqref{eq:constreqgen}, which can therefore be rewritten as
\begin{equation}
 M_L^2K=V_{,\phi}\,.
\end{equation}
This makes the limiting curvature behavior even more explicit: a bounded function $V_{,\phi}$ prevents singularities in the mean curvature $K$, and the limiting curvature scale is given by $M_L$.

In the fluid picture, the sound speed of the cuscuton field follows
\begin{equation}
 c_\mathrm{s,cusc}^2=\frac{p_{,X}}{\rho_{,X}}=\frac{P_{,X}}{P_{,X}+2XP_{,XX}}\,,
\end{equation}
which goes to infinity when $P_{,X}+2XP_{,XX}=0$. Therefore, the cuscuton is viewed as an incompressible fluid. An infinite propagation speed may appear unphysical at first, but we recall the second requirement for a cuscuton field: the scalar perturbations must not propagate. Let us consider the case of scalar perturbations about an FLRW background in the unitary gauge, so the perturbed metric is
\begin{equation}
 g_{\mu\nu}\mathrm{d}x^\mu\mathrm{d}x^\nu=-(1+2\Phi)\mathrm{d}t^2+2aB_{,i}\mathrm{d}x^i\mathrm{d}t+a^2(1-2\zeta)\delta_{ij}\mathrm{d}x^i\mathrm{d}x^j\,.
 \label{eq:metricug}
\end{equation}
The lapse and shift perturbations ($\Phi$ and $B$, respectively) can be eliminated from the Hamiltonian and momentum constraints, and the resulting second-order perturbed action for a generic $k$-essence scalar field is
\begin{equation}
 S^{(2)}_S=\int\mathrm{d}t\,\mathrm{d}^3\mathbf{x}\ a^3\left(\mathcal{G}_S\dot\zeta^2-\frac{\mathcal{F}_S}{a^2}(\vec{\nabla}\zeta)^2\right)\,,
\end{equation}
with
\begin{equation}
 \mathcal{G}_S=\frac{X}{H^2}(P_{,X}+2XP_{,XX})\,,\qquad\mathcal{F}_S=\epsilon \mpl^2\,,
\end{equation}
where $\epsilon\equiv -\dot H/H^2$. For a cuscuton scalar field with $P_{,X}+2XP_{,XX}=0$, it immediately follows that\footnote{One might worry about what happens if $H=0$ at some point in time (that would precisely correspond to the bounce point in a non-singular cosmology). Indeed, $\mathcal{G}_S$ would appear indefinite then. This issue is revisited in Sec.~\ref{sec:cosmopert}.} $\mathcal{G}_S=0$, hence the kinetic term for $\zeta$ drops out of the perturbed action and no scalar degree of freedom propagates.
This is precisely the desired behavior for the cuscuton field.
In particular, it implies that the infinite value for $c_\mathrm{s,cusc}^2=\mathcal{F}_S/\mathcal{G}_S$ is not physical since no information is carried by the scalar perturbations.

\subsection{Cuscuton bounce}\label{sec:cuscutonbounce}

As shown in the previous subsection, it is clear that cuscuton gravity is a limiting curvature theory with the prospect of avoiding singularities.
Let us show how this can be done explicitly. Let us start with the cuscuton gravity action together with additional matter:
\begin{equation}
 S=\int\mathrm{d}^4x\,\sqrt{-g}\left(\frac{\mpl^2}{2}R\pm M_L^2\sqrt{2X}-V(\phi)\right)+S^{\mathrm{(m)}}\,.
\end{equation}
In an FLRW background, the energy density and pressure of the cuscuton field reduce to $V(\phi)$ and $\pm M_L^2|\dot\phi|-V(\phi)$, respectively. Let $T_{\mu\nu}\equiv -(2/\sqrt{-g})\delta S^{\mathrm{(m)}}/\delta g^{\mu\nu}$ be the energy-momentum tensor of the additional matter, and let it be that of a perfect fluid, $T^\mu{}_\nu=\mathrm{diag}(-\rho,p\,\delta^i{}_j)$, in an FLRW background. Putting everything together, the resulting Friedmann equations are
\begin{align}
 3\mpl^2H^2&=\rho+V(\phi)\,, \nonumber \\
 2\mpl^2\dot H&=-(\rho+p)\mp M_L^2|\dot\phi|\,,
\end{align}
together with the usual conservation equation for matter, $\dot\rho+3H(\rho+p)=0$, and the constraint equation \eqref{eq:constreqFLRW} for the cuscuton field.
As we can see from the second Friedmann equation, there exists a regime where the effective null energy condition (NEC) can be violated, i.e., where $\dot H>0$. This is possible when the negative sign is chosen for the cuscuton Lagrangian, i.e., $\mathcal{L}=-M_L^2\sqrt{2X}-V(\phi)$, in which case the second Friedmann equation becomes $2\mpl^2\dot H=-(\rho+p)+M_L^2|\dot\phi|$. Then, $\dot H>0$ is possible when $M_L^2|\dot\phi|>\rho+p$, which can be achieved with the additional matter satisfying the usual NEC with $\rho+p\geq 0$. Therefore, several potential functions $V(\phi)$ can be derived giving rise to a non-singular bouncing solution. When $\rho+p>M_L^2|\dot\phi|$, the universe can be contracting in the past ($H<0$, $\dot H<0$) and expanding in the future ($H>0$, $\dot H<0$), and in the regime $\rho+p<M_L^2|\dot\phi|$ the universe undergoes a transition where $H$ passes through 0 and changes sign while $\dot H>0$. An example of such a background solution can be found in Ref.~\cite{Boruah:2018pvq}.

In what follows, let us consider the simplified case where the additional matter is described by a massless scalar field $\chi$, i.e.,
\begin{equation}
 S=\int\mathrm{d}^4x\,\sqrt{-g}\left(\frac{\mpl^2}{2}R-M_L^2\sqrt{2X}-V(\phi)-\frac{1}{2}g^{\mu\nu}\partial_\mu\chi\partial_\nu\chi\right)\,.
 \label{eq:actioncuscutonscalarfield}
\end{equation}
Then, the set of independent equations of motion with cuscuton constraint equation can be written as
\begin{align}
 3\mpl^2H^2-\left(\frac{1}{2}\dot\chi^2+V(\phi)\right)&=0\,, \label{Friedconstraint}\\
 \ddot\chi+3H\dot\chi&=0\,, \label{eq:chiEOM}\\
 \mathrm{sgn}(\dot\phi)3M_L^2H-V_{,\phi}&=0\,. \label{eq:cusconstreqFLRW2}
\end{align}
The matter ($\chi$) equation of motion can immediately be integrated to yield
\begin{equation}
 \dot\chi=\frac{\dot\chi_\mathrm{ini}}{a^3}\,,
\end{equation}
where the integration constant $\dot\chi_\mathrm{ini}\neq 0$ is set by the initial conditions
(we ignore the trivial solution with $\dot\chi_\mathrm{ini}=0$).
Taking a time derivative of the cuscuton constraint equation, one finds
\begin{equation}
 3M_L^2\dot H=V_{,\phi\phi}|\dot\phi|\,.
 \label{eq:Vppbkg}
\end{equation}
Therefore, in the bouncing phase where $\dot H>0$, one must have\footnote{Similarly, outside the bounce phase the NEC is satisfied if $V_{,\phi\phi}<0$. The point where $V_{,\phi\phi}=0$ defines the onset of NEC violation at which point $\dot H=0$. Consequently, note that $\dot\phi\neq 0$ at all times. This avoids possible issues with the point $X=0$ in the cuscuton action.} $V_{,\phi\phi}>0$.
Let us perform a Taylor expansion of the potential function about the bouncing point:
\begin{equation}
 V(\phi)=V_0+\frac{1}{2}m^2\phi^2+\mathcal{O}(\phi^3)\,.
\end{equation}
We arbitrarily choose $\phi=0$ to correspond to the bouncing point where $H=0$.
We note that there is no $\mathcal{O}(\phi)$ term since $V_{,\phi}=0$ when $H=0$ according to Eq.~\eqref{eq:cusconstreqFLRW2}. Also, we defined $m^2\equiv V_{,\phi\phi}(\phi=0)>0$. Substituting the Taylor expansion for $V(\phi)$ into the constraint equation \eqref{eq:cusconstreqFLRW2}, one can write
\begin{equation}
 \phi=3\,\mathrm{sgn}(\dot\phi)\frac{M_L^2}{m^2}H+\mathcal{O}(H^2)\,,
\end{equation}
which is also expanded about the bouncing point $H=0$.
Accordingly, the Friedmann equation up to $\mathcal{O}(H^3)$ can be written as\footnote{At this point, let us point out that if there were additional matter sources, such as matter ($\rho^{(\mathrm{m})}\propto a^{-3}$) and radiation ($\rho^{(\mathrm{rad})}\propto a^{-4}$), they would have subdominant energy densities in the bounce phase, so they can be safely ignored.
However, one might worry about anisotropies, whose energy density scales as $a^{-6}$, just as the massless scalar field.
Therefore, an underlying assumption is that, as an initial condition, anisotropies are tuned to have a smaller energy density than the massless scalar field.
Otherwise, the anisotropies could disrupt the bounce.
If such a tuning is not performed, one would rather have to invoke an `isotropization' process that effectively washes out anisotropies in the contracting phase.
This remains an issue of ongoing research (see, e.g., Refs.~\cite{BKL,Lin:2017fec}).}
\begin{equation}
 3\mplt^2H^2\simeq\frac{1}{2}\dot\chi^2+V_0\,,
\end{equation}
where we define an effective Planck mass
\begin{equation}
 \mplt^2\equiv \mpl^2\left(1-\frac{3}{2}\frac{M_L^4}{m^2\mpl^2}\right)\,,
 \label{eq:defmplt}
\end{equation}
and $V_0$ acts as an effective cosmological constant. Since $\dot\chi^2/2=\dot\chi_\mathrm{ini}^2/2a^6>0$, it must be that $V_0<0$ when $H=0$.
Similarly, the second Friedmann equation can be written as
\begin{equation}
 2\mplt^2\dot H\simeq -\dot\chi^2\,.
 \label{eq:dotHeqmplt}
\end{equation}
This makes explicit that one needs a negative gravitational coupling in order to have a bouncing phase with $\dot H>0$. This is possible when $\mplt^2<0$ or equivalently
\begin{equation}
 m^2<\frac{3}{2}\frac{M_L^4}{\mpl^2}\,.
\end{equation}
Writing the resulting solution about the bounce point as a Taylor series\footnote{We also arbitrarily set the bounce point at $t=0$.}
$a(t)=a_0+\ddot a_0t^2/2+\mathcal{O}(t^3)$, the Friedmann equations can finally be solved to find
\begin{equation}
 a(t)=\bigg(\frac{\dot\chi_\mathrm{ini}}{\sqrt{2|V_0|}}\bigg)^{1/3}\left(1+\frac{1}{2}\frac{V_0}{\mplt^2}t^2+\mathcal{O}(t^3)\right)\,,\qquad H(t)=\frac{V_0}{\mplt^2}t+\mathcal{O}(t^2)\,,\qquad\dot H=\frac{V_0}{\mplt^2}+\mathcal{O}(t)\,.
 \label{eq:bouncebckgrndseries}
\end{equation}

\section{Revisiting cosmological perturbations in the cuscuton bounce}\label{sec:cosmopert}

Cosmological perturbations in cuscuton gravity have already been explored in Refs.~\cite{Afshordi:2007yx,Boruah:2017tvg,Boruah:2018pvq}. Let us revisit the issue of the perturbations in a bouncing phase, highlighting some important subtleties.

\subsection{Scalar curvature perturbations: spatially-flat gauge as an example}\label{sec:sfg}

Let us consider scalar perturbations with the following perturbed metric:
\begin{equation}
 g_{\mu\nu}\mathrm{d}x^\mu\mathrm{d}x^\nu=-(1+2\Phi)\mathrm{d}t^2+2aB_{,i}\mathrm{d}x^i\mathrm{d}t+a^2[(1-2\Psi)\delta_{ij}+2E_{,ij}]\mathrm{d}x^i\mathrm{d}x^j\,.
\end{equation}
Furthermore, the fields are generally perturbed as
\begin{equation}
 \phi(t,\mathbf{x})=\phi(t)+\delta\phi(t,\mathbf{x})\,,\qquad\chi(t,\mathbf{x})=\chi(t)+\delta\chi(t,\mathbf{x})\,.
\end{equation}
The gauge transformations with respect to the infinitesimal diffeomorphism $x^\mu\rightarrow x^\mu-\xi^\mu$ can be written as
\begin{equation}
 \delta_{\bm{\xi}}\Phi=\dot\xi^0\,,\quad\delta_{\bm{\xi}}\Psi=-H\xi^0\,,\quad\delta_{\bm{\xi}} B=a^2\dot\xi-\frac{1}{a}\xi^0\,,\quad\delta_{\bm{\xi}} E=\xi\,,\quad\delta_{\bm{\xi}}(\delta\phi)=\xi^0\dot\phi\,,\quad\delta_{\bm{\xi}}(\delta\chi)=\xi^0\dot\chi\,,
\end{equation}
where $\xi^0$ and $\xi$ are defined such that
\begin{equation}
 \bm{\xi}\equiv\xi_\mu\mathbf{d}x^\mu=-\xi^0\mathbf{d}t+a^2\xi_{,i}\mathbf{d}x^i\,.
\end{equation}

Let us start by exploring the spatially-flat gauge, where $\Psi=E=0$, so the second-order action is to be described by three non-dynamical variables $\Phi$, $B$, $\delta\phi$ and one dynamical variable $\delta\chi$.
Each perturbation variable in the spatially-flat gauge has its corresponding gauge-invariant definition,
\begin{align}
 \Phi^S&\equiv\Phi+\partial_t\left(\frac{\Psi}{H}\right)\,,\nonumber\\
 B^S&\equiv B-a^2\dot E-\frac{1}{aH}\Psi\,,\nonumber\\
 \delta\phi^S&\equiv\delta\phi+\frac{\dot\phi}{H}\Psi\,,\nonumber\\
 \delta\chi^S&\equiv\delta\chi+\frac{\dot\chi}{H}\Psi\,,
\end{align}
where the superscript $S$ denotes the gauge-invariant variables in the spatially-flat gauge.
Since $\Phi^S$, $B^S$, and $\delta\phi^S$ are non-dynamical, one can derive the following constraint equations by varying the action \eqref{eq:actioncuscutonscalarfield} (expanded to second order in perturbations and transformed to Fourier space) with respect to those non-dynamical variables,
\begin{align}
 \Phi^S_k=&~\frac{\mpl^2\dot\chi}{\Xi}\left(F_1H\delta\chi^S_k+F_2\delta\dot\chi^S_k\right)\,,\nonumber\\
 B^S_k=&~\frac{a\mpl^2\dot\chi}{k^2\Xi}\left\{\left[3F_1H^2-2\tilde Y\left(\frac{k^2}{a^2}-3\dot H\right)\right]\delta\chi^S_k+F_1H\delta\dot\chi^S_k\right\}\,,\nonumber\\
 \delta\phi^S_k=&~\frac{2\mpl^4F_2\dot\chi}{M_L^2\Xi}\left(\tilde Y\delta\chi^S_k-H\delta\dot\chi^S_k\right)\,, \label{eq:constreqssfg}
\end{align}
where we defined
\begin{align}
 \Xi\equiv&~2\mpl^4(F_1H^2+F_2\tilde{Y})\,,\qquad\tilde{Y}\equiv\frac{\dot\chi^2}{2\mpl^2}\,,\nonumber\\
 F_1\equiv&~2[(k/a)^2+3\tilde{Y}]\,,\qquad F_2\equiv 2(\dot H+\tilde{Y})\,.
\end{align}
Using these constraints, the second-order perturbed action reduces to
\begin{equation}
 S_S^{(2)}=\mpl^2\int\mathrm{d}t\,\mathrm{d}^3\mathbf{k}~a\left(\frac{H}{\dot\chi}\right)^2z^2\left[\left(\delta\dot\chi^S_k\right)^2-c_S^2\frac{k^2}{a^2}\left(\delta\chi^S_k\right)^2\right]\,,
 \label{eq:SS2S1}
\end{equation}
where
\begin{equation}
 z^2\equiv\frac{2a^2\mpl^2F_1\dot\chi^2}{\Xi}\,,
 \label{eq:z2gen}
\end{equation}
which can also be written as $z^2=2a^2\mathcal{A}$ with
\begin{equation}
 \mathcal{A}\equiv\frac{\tilde{Y}[(k/a)^2+3\tilde{Y}]}{(k/a)^2H^2+\tilde{Y}(3H^2+\dot H+\tilde{Y})}\,.
\end{equation}
This is the same expression as found by Ref.~\cite{Boruah:2018pvq} or by Ref.~\cite{Iyonaga:2018vnu} when reduced to the minimal cuscuton model in Eq.~\eqref{eq:actioncuscutonscalarfield}.
According to the second Friedmann equation, we have
\begin{equation}
 \dot H+\tilde Y=\frac{M_L^2}{2\mpl^2}|\dot\phi|>0\,.
\end{equation}
It becomes apparent that $\mathcal{A}>0$, and thus $z^2>0$ at any point in time. Therefore, the theory appears to have no ghost instability.
The sound speed squared in Eq.~\eqref{eq:SS2S1} is given by
\begin{equation}
 c_S^2=\frac{H^2(k/a)^4+A_2(k/a)^2+A_0}{H^2(k/a)^4+B_2(k/a)^2+B_0}\,,
 \label{eq:cS2gen}
\end{equation}
where we define
\begin{align}
 A_2\equiv&~\tilde{Y}(12H^2+3\dot H+\tilde{Y})+2\dot H^2-H\ddot H\,,\nonumber\\
 A_0\equiv&~\tilde{Y}^2(15H^2+\dot H-\tilde{Y})-\tilde{Y}\left(12H^2\dot H-2\dot H^2+3H\ddot H\right)\,,\nonumber\\
 B_2\equiv&~\tilde{Y}(6H^2+\dot H+\tilde{Y})\,,\nonumber\\
 B_0\equiv&~3\tilde{Y}^2(3H^2+\dot H+\tilde{Y})\,.
\end{align}
Clearly, we find $c_S\rightarrow 1$ in the UV limit $k/a\rightarrow\infty$, so there is no gradient instability as well.
There is a slight caveat when $H=0$, but this will be explained below.

Finally, to normalize the action \eqref{eq:SS2S1}, we define the curvature perturbation by
\begin{equation}
 \zeta\equiv-\frac{H}{\dot\chi}\delta\chi^S\,.
 \label{eq:zetasfg}
\end{equation}
This allows us to write the perturbed action as
\begin{equation}
 S_S^{(2)}=\mpl^2\int\mathrm{d}t\,\mathrm{d}^3\mathbf{k}~az^2\left(\dot\zeta_k^2-c_S^2\frac{k^2}{a^2}\zeta_k^2\right)\,,
 \label{eq:SS2sfgf}
\end{equation}
The corresponding equation of motion is
\begin{equation}
 \ddot\zeta_k+\left(H+2\frac{\dot z}{z}\right)\dot\zeta_k+c_S^2\frac{k^2}{a^2}\zeta_k=0\,.
 \label{eq:zetaEOM}
\end{equation}
One can check that every coefficient is regular when $H=0$ (from here on, we call this point `Hubble crossing'), which corresponds to the bounce point in a non-singular bounce phase.
However, as we will see in the next sub-section, the spatially-flat gauge is actually ill defined at that point in time.

\subsection{(In)validity of the spatially-flat gauge at Hubble crossing}

While cosmological perturbations appear to be well behaved throughout cosmic time, in particular through a non-singular bounce, it turns out that the results in the spatially-flat gauge cannot be trusted at Hubble crossing, i.e., $H=0$ crossing.
Indeed, writing the constraints equations \eqref{eq:constreqssfg} in terms of the curvature perturbation, defined in Eq.~\eqref{eq:zetasfg}, and expanding about $H=0$, one finds,
assuming $\dot\phi>0$ without loss of generality,
\begin{equation}
 \Phi^S_k=-\partial_t\left(\frac{\zeta_k}{H}\right)+\mathcal{O}(H^0)\,,\qquad B^S_k=\frac{1}{aH}\zeta_k+\mathcal{O}(H^0)\,,\qquad\delta\chi^S_k=-\frac{\dot\chi}{H}\zeta_k\,,\qquad\delta\phi^S_k=-\frac{\dot\phi}{H}\zeta_k+\mathcal{O}(H^0)\,,
\end{equation}
and so every perturbation variable diverges as $H\rightarrow 0$.
Consequently, the spatially-flat gauge is actually not well defined at that point in time, and the conclusions about stability throughout a non-singular bouncing phase might not be trustable anymore. Fortunately, we find below that other gauges are well defined at Hubble crossing and confirm the results about stability.

\subsection{Going to another gauge: comoving gauge with respect to the cuscuton field}

The comoving gauge with respect to $\phi$ is defined by $E=\delta\phi=0$. This is the gauge used in Ref.~\cite{Iyonaga:2018vnu}. The gauge-invariant variables in this gauge are defined by
\begin{equation}
 \Phi^\phi\equiv\Phi-\partial_t\left(\frac{\delta\phi}{\dot\phi}\right)\,,\qquad
 B^\phi\equiv B-a^2\dot E+\frac{1}{a}\frac{\delta\phi}{\dot\phi}\,,\qquad
 \Psi^\phi\equiv\Psi+H\frac{\delta\phi}{\dot\phi}\,,\qquad
 \delta\chi^\phi\equiv\delta\chi-\dot\chi\frac{\delta\phi}{\dot\phi}\,.
\end{equation}
Note that the superscript $\phi$ indicates a gauge-invariant variable in the comoving gauge with respect to the cuscuton field.
From these definitions, gauge-invariant variables in this comoving gauge can be represented in terms of the gauge-invariant variables in the spatially-flat gauge as shown below. Furthermore, upon using the constraint equations \eqref{eq:constreqssfg}, expressing the single dynamical degree of freedom in terms of the curvature perturbation $\zeta$, and expanding about Hubble crossing, one finds that the divergences found in the spatially-flat gauge cancel out exactly,
\begin{align}
 \Phi^{\phi}_k & =\Phi^{S}_k - \partial_{t}\left(\frac{\delta \phi^{S}_k}{\dot{\phi}}\right) = -\frac{4}{1+3\tilde Y(k/a)^{-2}}\zeta_k + {\cal O}(H)\,,\nonumber\\
 B^{\phi}_k & = B^{S}_k + \frac{1}{a \dot{\phi}} \delta \phi^{S}_k = - \frac{3\dot\chi^2}{M_L^2a\dot\phi}\left(\frac{a}{k}\right)^2 \dot{\zeta}_k + {\cal O}(H)\,,\nonumber \\
 \Psi^{\phi}_k & = H \frac{\delta \phi^{S}_k}{\dot{\phi}} = - \zeta_k + {\cal O}(H)\,,\nonumber\\
 \delta \chi^{\phi}_k &  = \delta \chi^{S}_k - \dot{\chi} \frac{\delta \phi^{S}_k}{\dot{\phi}} = - \frac{2\mpl^2\dot{\chi}}{M_L^2\dot{\phi}}\dot{\zeta}_k + {\cal O}(H)\,.
 \label{eq:Stophigauges}
\end{align}
The curvature perturbation variable $\zeta$ in this comoving gauge, as used in Ref.~\cite{Iyonaga:2018vnu}, is given by
\begin{align}
 \zeta \equiv - \Psi^{\phi} - \frac{H}{\dot\chi}\delta\chi^\phi = - \frac{H}{\dot\chi}\delta \chi^{S}\,,
\end{align}
so it is the same curvature perturbation variable as in the spatially-flat gauge [recall Eq.~\eqref{eq:zetasfg}].
Therefore, the perturbed action and equation of motion found in Sec.~\ref{sec:sfg} apply, and in particular, the stability results (no ghost and no gradient instabilities) translate to the comoving gauge. This time though, the gauge is well defined at Hubble crossing since divergences in the perturbation variables cancel out as $H\rightarrow 0$. Consequently, the results can be trusted at that point in time.
Note that this gauge would be ill defined, however, if there were a point in time where $\dot\phi=0$ or $\dot\chi=0$ (in which case $\tilde Y=0$). Such situations can occur, for instance, when a scalar field oscillates\footnote{This is a common issue in preheating, where the inflaton oscillates and the field velocity crosses zero periodically.
At such points in time, it is known that the comoving gauge with respect to the inflaton field is not well defined and other gauges must be used, such as the Newtonian gauge (see, e.g., Ref.~\cite{Finelli:1998bu}).}.
This is not an issue here since $\dot\chi\neq 0$ [though this could change with the addition of a potential term $U(\chi)$] and $\dot\phi\neq 0$ at all times.

The above result remains somewhat surprising. Indeed, if we started by expanding the action in the comoving gauge with respect to $\chi$, so with the metric \eqref{eq:metricug}, we would obtain the following constraint equations by varying the second-order expanded action with respect to the lapse and shift perturbations, $\Phi$ and $B$, respectively:
\begin{align}
 (\tilde Y-3H^2)\Phi_k+\frac{H}{\mpl}\left(\frac{k}{a}\right)^2B_k+3H\dot\zeta_k+\left(\frac{k}{a}\right)^2\zeta_k-\frac{\dot\chi}{\mpl^2}\dot{\delta\chi_k}&=0\,,\nonumber\\
 H\Phi_k-\dot\zeta_k-\frac{\dot\chi}{2\mpl^2}\delta\chi_k&=0\,.
\end{align}
To eliminate $\Phi$ and $B$ from the perturbed action, one solves the above two equations for the lapse and shift perturbations, but in doing so one must be able to divide by $H$, which would be an issue when $H=0$.
However, as shown above, the other perturbations $\zeta$ and $\delta\chi$ scale precisely the right away to leading order in small $H$ to cancel out the would-be divergences when $H=0$. Therefore, the solutions to $\Phi$, $B$, and consequently, the reduced perturbed action all have well-defined expressions even when $H=0$.

\subsection{Comoving gauge with respect to the matter field}

The comoving gauge with respect to $\chi$ is defined by $E=\delta\chi=0$. This is the gauge used in Ref.~\cite{Boruah:2018pvq}. The gauge-invariant variables in this gauge are defined by
\begin{equation}
 \Phi^\chi\equiv\Phi-\partial_t\left(\frac{\delta\chi}{\dot\chi}\right)\,,\qquad
 B^\chi\equiv B-a^2\dot E+\frac{1}{a}\frac{\delta\chi}{\dot\chi}\,,\qquad
 \Psi^\chi\equiv\Psi+H\frac{\delta\chi}{\dot\chi}\,,\qquad
 \delta\phi^\chi\equiv\delta\phi-\dot\phi\frac{\delta\chi}{\dot\chi}\,.
\end{equation}
Note that the superscript $\chi$ indicates a gauge-invariant variable in the comoving gauge with respect to the matter field.
Using the same methodology as in the previous sub-section with the other comoving gauge, the gauge-invariant variables in this comoving gauge can be represented in terms of the gauge-invariant variables in the spatially-flat gauge as shown below. Again, expanding about Hubble crossing one finds that the divergences found in the spatially-flat gauge cancel out exactly:
\begin{align}
 \Phi^{\chi}_k &= \Phi^{S}_k - \partial_{t}\left(\frac{\delta \chi^{S}_k}{\dot{\chi}}\right) = - \frac{2\mpl^2}{\dot\chi^2}\left(\frac{k}{a}\right)^2 \zeta_k  + {\cal O}(H)\,,\nonumber\\
 B^{\chi}_k &= B^{S}_k + \frac{1}{a}\frac{\delta\chi^{S}_k}{\dot{\chi}} = -\frac{2\mpl^2}{M_L^2a\dot\phi}\left[1+3\tilde Y\left(\frac{a}{k}\right)^2\right]\dot{\zeta}_k + {\cal O}(H)\,,\nonumber\\
 \Psi^{\chi}_k &= H \frac{\delta \chi^{S}_k}{\dot{\chi}} = - \zeta_k\,,\nonumber\\
 \delta \phi^{\chi}_k &= \delta \phi^{S}_k - \dot{\phi} \frac{\delta \chi^{S}_k}{\dot{\chi}} = \frac{2 \mpl^2}{M_L^2} \dot{\zeta}_k + {\cal O}(H)\,.
 \label{eq:Stochigauges}
\end{align}
The curvature perturbation variable $\zeta$ in this comoving gauge, as used in Ref.~\cite{Boruah:2018pvq}, is given by
\begin{align}
 \zeta \equiv - \Psi^{\chi} = - \frac{H}{\dot\chi}\delta \chi^{S}\,,
\end{align}
so it is the same curvature perturbation variable as in the spatially-flat gauge.
Consequently, the same results follow: the comoving gauge with respect to the matter field is well defined at Hubble crossing, and the results about the stability of the perturbations throughout a non-singular bounce phase are valid.

\subsection{Newtonian gauge}

\subsubsection{Validity at Hubble crossing}

The Newtonian gauge is given by $B=E=0$. The gauge-invariant perturbation variables are then defined by
\begin{equation}
 \Phi^N\equiv\Phi+\partial_t[a(B-a^2\dot E)]\,,\quad
 \Psi^N\equiv\Psi-aH(B-a^2\dot E)\,,\quad
 \delta\phi^N\equiv\delta\phi+a\dot\phi(B-a^2\dot E)\,,\quad
 \delta\chi^N\equiv\delta\chi+a\dot\chi(B-a^2\dot E)\,,
\end{equation}
where the superscript $N$ denotes the Newtonian gauge.
With the same methodology as before, relating to the spatially-flat gauge and expanding about Hubble crossing, we find
\begin{align}
 \Phi^{N}_k &= \Phi^{S}_k + \partial_{t} (a B^{S}_k)
 =\frac{\tilde Y[-(k^2/a^2)(\tilde Y+\dot H)\zeta_k+H(k^2/a^2+3\tilde Y)\dot\zeta_k]}{\mathcal{D}}
 =-\zeta_k+\mathcal{O}(H)\,,\nonumber\\
 \Psi^{N}_k &= - a H B^{S}_k
 =\frac{\tilde Y[-(k^2/a^2)(\tilde Y+\dot H)\zeta_k+H(k^2/a^2+3\tilde Y)\dot\zeta_k]}{\mathcal{D}}
 =-\zeta_k+\mathcal{O}(H)\,,\nonumber\\
 \delta\phi^{N}_k &= \delta \phi^{S}_k + a \dot{\phi} B^{S}_k
 =-\frac{6\mpl^2}{M_L^2}\frac{\tilde Y(\tilde Y+\dot H)[(k^2/a^2)H\zeta_k+\tilde Y\dot\zeta_k]}{\mathcal{D}}
 =-\frac{3\dot\chi^2}{M_L^2}\left(\frac{a}{k}\right)^2\dot\zeta_k+\mathcal{O}(H)\,,\nonumber\\
 \delta\chi^{N}_k &= \delta \chi^{S}_k + a \dot{\chi} B^{S}_k
 =-\mpl\sqrt{2\tilde Y}\frac{(k^2/a^2+3\tilde Y)[(k^2/a^2)H\zeta_k+\tilde Y\dot\zeta_k]}{\mathcal{D}}
 =-\frac{2\mpl^2\dot\chi}{M_L^2|\dot\phi|}\left[1+3\tilde Y\left(\frac{a}{k}\right)^2\right]\dot\zeta_k+\mathcal{O}(H)\,,
 \label{eq:Ngtransfs}
\end{align}
where we define $\mathcal{D}\equiv (k^2/a^2)[H^2(k^2/a^2+3\tilde Y)+\tilde Y(\tilde Y+\dot H)]$ for convenience.
Once again, we conclude that the Newtonian gauge is well behaved at Hubble crossing. However, it turns out that the Newtonian gauge can appear ill behaved at other points in time, as we will see in the next sub-section.

\subsubsection{General equations}

Let us derive the general equations in the Newtonian gauge from perturbing the equations of motion rather than working at the level of the action.
Starting from the cuscuton general constraint equation \eqref{eq:constreqgen}, perturbing to linear order yields
\begin{equation}
 \mathrm{sgn}(\dot{\phi})3(H\Phi+\dot\Psi)+|\dot{\phi}|^{-1}a^{-2}\nabla^2\delta\phi+M_L^{-2}V_{,\phi\phi}\delta\phi=0\,.
 \label{eq:pertcuscuEOM1}
\end{equation}
Recalling the expression \eqref{eq:Vppbkg} for $V_{,\phi\phi}$, the perturbed constraint reduces to
\begin{equation}
 3\dot{\phi}(H\Phi+\dot\Psi)+a^{-2}\nabla^2\delta\phi+3\dot H\delta\phi=0\,,
\end{equation}
and upon transforming to Fourier space, one can isolate the perturbed cuscuton field as follows:
\begin{equation}
 \delta\phi_k=\frac{3\dot{\phi}(H\Phi_k+\dot\Psi_k)}{k^2/a^2-3\dot H}\,,
 \label{eq:constraintdeltaphiNg}
\end{equation}
as long as $k^2/a^2\neq 3\dot H$. Outside the bounce phase, $\dot H\leq 0$ and one can always isolate the perturbation of the cuscuton field as above. However, in the bouncing phase where the NEC is violated ($\dot H>0$), there is a point in time where $k^2/a^2=3\dot H$ for a range of infrared modes, $0<k\leq k_\mathrm{max}\equiv a_B\sqrt{3\dot H_B}$, where $a_B$ and $\dot H_B$ are the values of the scale factor and time derivative of the Hubble parameter at the bounce point (Hubble-crossing time).
At these points in time, the cuscuton field fluctuation cannot be eliminated from the above equation as otherwise divergences would appear in the perturbation equations of motion.

The perturbed Einstein equations in the Newtonian gauge for the model of cuscuton gravity with a massless scalar field can be found as follows:
\begin{align}
 3H(\dot\Phi_k+H\Phi_k)+\frac{k^2}{a^2}\Phi_k=&~-\frac{1}{2\mpl^2}\left(\dot\chi(\dot{\delta\chi}_k-\dot\chi\Phi_k)+V_{,\phi}\delta\phi_k\right)\,, \nonumber \\
 \dot\Phi_k+H\Phi_k=&~\frac{1}{2\mpl^2}\left(\dot\chi\delta\chi_k-\mathrm{sgn}(\dot\phi)M_L^2\delta\phi_k\right)\,, \nonumber \\
 \ddot\Phi_k+4H\dot\Phi_k+(3H^2+2\dot H)\Phi_k=&~\frac{1}{2\mpl^2}\left(\dot\chi(\dot{\delta\chi}_k-\dot\chi\Phi_k)+M_L^2(|\dot\phi|\Phi_k-\mathrm{sgn}(\dot\phi)\dot{\delta\phi}_k)-V_{,\phi}\delta\phi_k\right)\,,
 \label{eq:allNgpertEEs}
\end{align}
where the fact that $\Psi_k=\Phi_k$ is used.
Using the constraint equation for $\delta\phi_k$, Eq.~\eqref{eq:constraintdeltaphiNg}, and combining the above, one finds
\begin{equation}
 (1+\mathcal{C}_2)\ddot\Phi_k+(\mathcal{C}_1+7H(1+\mathcal{C}_2)+\mathcal{C}_3)\dot\Phi_k+\left(\frac{k^2}{a^2}+(6H^2+\dot H)(1+\mathcal{C}_2)+H(\mathcal{C}_1+\mathcal{C}_3)-\tilde Y\right)\Phi_k=0\,,
\end{equation}
where
\begin{equation}
 \mathcal{C}_1\equiv 3\frac{\ddot H-6H\tilde Y}{k^2/a^2-3\dot H}\,,\qquad \mathcal{C}_2\equiv 3\frac{\dot H+\tilde Y}{k^2/a^2-3\dot H}\,,\qquad \mathcal{C}_3\equiv\mathcal{C}_2\frac{2Hk^2/a^2+3\ddot H}{k^2/a^2-3\dot H}\,.
\end{equation}
The equation of motion for $\Phi_k$ thus appears singular only when $k^2/a^2=3\dot H$. However, as we will see below, it turns out that $\Phi_k$ remains well defined even at that point in time.

We recall that the gauge-invariant perturbation variables in the Newtonian gauge can be written in terms of the curvature perturbation according to Eq.~\eqref{eq:Ngtransfs}.
By inspection, it is already obvious that every perturbation variable in the Newtonian gauge remains finite and well defined when $k^2/a^2=3\dot H$.
In fact, one can check that the common denominator $\mathcal{D}$ is always positive definite.
The important thing to notice is that the combination $H\Phi_k+\dot\Psi_k$ scales as $k^2/a^2-3\dot H$, thus canceling the would-be divergence in the perturbations.
Indeed, one can check that
\begin{equation}
 H\Phi_k+\dot\Psi_k=-\frac{\tilde Y}{\mathcal{D}}\left(\frac{k^2}{a^2}-3\dot H\right)\left(H\frac{k^2}{a^2}\zeta_k+\tilde Y\dot\zeta_k\right)\,.
\end{equation}
As a consistency check, plugging the above in the constraint equation \eqref{eq:constraintdeltaphiNg} confirms that the apparent singularity at $k^2/a^2=3\dot H$ disappears and one recovers the Newtonian gauge expression for $\delta\phi_k$ as a function of the curvature perturbation as found in Eq.~\eqref{eq:Ngtransfs}.

Conversely, the curvature perturbation can be written in terms of Newtonian gauge quantities as follows: $\zeta_k=-\Psi_k-(H/\dot\chi)\delta\chi_k$. Using the second equation in \eqref{eq:allNgpertEEs} to replace $\delta\chi_k$ and the constraint equation \eqref{eq:constraintdeltaphiNg} to replace $\delta\phi_k$, one finds
\begin{equation}
 \zeta_k=-\Phi_k-\frac{H}{\tilde Y}\frac{k^2/a^2+3\tilde Y}{k^2/a^2-3\dot H}(\dot\Phi_k+H\Phi_k)\,.
\end{equation}
As expected, there appears to be a singularity in the transformation when $k^2/a^2=3\dot H$, but again, this is not the case since the combination $\dot\Phi_k+H\Phi_k$ is proportional to $k^2/a^2-3\dot H$.

In summary, the Newtonian gauge is well defined everywhere in time.
In particular, there is no issue appearing around Hubble crossing.
This is made explicit from the full equation of motion for the Newtonian potential.
The only crossing time that needs special care is when a certain $k$ mode satisfies $k=a\sqrt{3\dot H}$.
At that point, the equation of motion for $\Phi_k$ appears to diverge, and so, the Newtonian gauge might not be best suited for analyzing the perturbations around that point.
Nevertheless, we confirm that the perturbations must remain finite when $k^2/a^2=3\dot H$.
This is made explicit when evaluating the Newtonian gauge perturbation variables in terms of the curvature perturbation, which is well defined when $k^2/a^2=3\dot H$.
Still, in the vicinity of the point where $k^2/a^2=3\dot H$, it might be better to work in another gauge, e.g., one of the comoving gauges.

\subsection{More comments on stability}

Since both comoving gauges are well-behaved throughout the bounce phase, in particular about $H=0$, it appears convenient to use either to study the perturbations more closely in the bounce phase.
However, one has to be slightly careful with certain limits.
For example, taking the UV limit of action \eqref{eq:SS2sfgf} yields
\begin{equation}
 S_S^{(2)}\simeq 2\mpl^2\int\mathrm{d}t\,\mathrm{d}^3\mathbf{k}~a^3\frac{\tilde Y}{H^2}\left(\dot\zeta_k^2-\frac{k^2}{a^2}\zeta_k^2\right)\,.
\end{equation}
Such a UV limit confirms that perturbations are stable for small wavelength fluctuations.
This was already found by Ref.~\cite{Lin:2017fec}, who studied an effective field theory action which was that of cuscuton gravity in the unitary gauge in the scalar sector.
However, the above UV limit appears to be divergent when $H=0$.
This is due to the fact that the UV limit is actually not the above when the Hubble parameter crosses zero.
Setting $H=0$ first, one can perform the UV limit again to find
\begin{equation}
 S_S^{(2)}\simeq 4\frac{\mpl^4}{M_L^2}\int\mathrm{d}t\,\mathrm{d}^3\mathbf{k}~\frac{ak^2}{|\dot\phi|}\left(\dot\zeta_k^2-\left(1+\frac{2\dot H}{\tilde Y}\right)\frac{k^2}{a^2}\zeta_k^2\right)\,.
\end{equation}
Using Eq.~\eqref{eq:dotHeqmplt}, which is a background equation expanded about $H=0$, the sound speed squared can be simplified to
\begin{equation}
 c_S^2\simeq 1+\frac{4m^2\mpl^2}{3M_L^4-2m^2\mpl^2}\,,
 \label{eq:cs2max}
\end{equation}
where we recall $0<m^2\mpl^2/M_L^4<3/2$, $m^2$ being the second derivative of the cuscuton potential at the bounce point.
Consequently, one finds that perturbations in the limit $k\rightarrow\infty$ propagate at a superluminal speed at the bounce point.
While this is an interesting result, it is unlikely that it can lead to any observational test.
Indeed, a large $k$ mode would see its sound speed grow superluminal only very close to the bounce point for a very short time interval.
Also, very small wavelength fluctuation modes are generally difficult to observe (in the CMB or LSS), unless, in this context, they are stretched after the bounce, e.g., through a period of inflation as in models of bounce-inflation (see, e.g., Ref.~\cite{Qiu:2015nha}).
Let us also mention that superluminality does not represent an issue with respect to causality here \cite{superluminal}.
For a fixed wavenumber and at a given time, the above tells us that the light cone for $\zeta$ fluctuations is redefined by a sound speed that is at most the above value of $c_S$.
Indeed, one can check that Eq.~\eqref{eq:cs2max} serves as a finite upper bound on $c_S^2$ for any value of $k$ and at all times.

Let us have a closer look at the general expression for the sound speed [given in Eq.~\eqref{eq:cS2gen}] in different regimes.
We already commented that in the UV limit, $k\rightarrow\infty$, one finds $c_S^2\rightarrow 1$, except at the very point where $H=0$.
This is true in general when $k/a\gg\mathcal{O}(\sqrt{\tilde Y})$.
In the opposite regime, when $k/a\ll\mathcal{O}(\sqrt{\tilde Y})$ and in particular in the IR limit $k\rightarrow 0$, one finds $c_S^2\simeq A_0/B_0$.
In the bounce phase, and in particular, close to Hubble crossing, this can be approximated as follows:
\begin{equation}
 c_S^2\simeq -\frac{1}{3}+\frac{2\dot H}{3\tilde Y}\simeq -\frac{1}{3}+\frac{4m^2\mpl^2}{3(3M_L^4-2m^2\mpl^2)}\,.
\end{equation}
We thus see that requiring $0<c_S^2\leq 1$ amounts to constraining the parameter $m$ such that
\begin{equation}
 \frac{1}{2}<\frac{m^2\mpl^2}{M_L^4}\leq 1\,,
 \label{eq:boundm2}
\end{equation}
and in particular, one can remain far from the regime $c_S^2\ll 1$ as long as $m^2\approx M_L^4/\mpl^2$.
The condition $c_S^2>0$ is usually necessary to avoid gradient instabilities in the UV.
In the IR (in fact, throughout the spectrum), it is still relevant to ensure the classical well-posedness of the partial differential equations, i.e., to ensure strong hyperbolicity (e.g., see Ref.~\cite{Ijjas:2018cdm}).
Furthermore, $c_S^2\sim\mathcal{O}(1)$ can help with regards to strong coupling issues that often arise in NEC-violating regimes, especially within $k$-essence theory \cite{deRham:2017aoj}.
A more detailed analysis of the strong coupling scale is left for future work, but the above indicates that it might be sufficiently high to validate the use of cuscuton gravity as an effective field theory through a non-singular bounce.

One is left with the intermediate regime, where $k/a\sim\mathcal{O}(\sqrt{\tilde Y})$.
In this regime, all terms contribute more or less equally in the expression \eqref{eq:cS2gen} for $c_S^2$, and in particular, one can say that $c_S^2\sim A_2/B_2$.
Close to the bounce point, this means $c_S^2\sim 1+2\dot H/\tilde Y$, which is the same expression as found above for the UV limit at the point where $H=0$. Hence, an additional approximation close to $H=0$ yields Eq.~\eqref{eq:cs2max}.
Then, the bound \eqref{eq:boundm2} implies $2\lesssim c_S^2\lesssim 5$.
Consequently, superluminality is unavoidable close to the bounce point even in the intermediate regime where $k/a\sim\mathcal{O}(\sqrt{\tilde Y})$.
Nevertheless, the sound speed remains bounded and of order unity [or at most $\mathcal{O}(10)$].
Therefore, strong coupling problems arising with superluminality\footnote{For classically stable non-singular cosmologies within Horndeski theory, superluminality is often unavoidable \cite{Dobre:2017pnt}. Within beyond-Horndeski theory, this can be avoided for specific models \cite{Mironov:2019mye}.} might not be too severe here since we do not find $c_S^2\gg 1$, although a proper analysis of this issue remains to be completed.

Let us end by noting that away from the bounce phase, it can be explicitly checked that $c_S^2\simeq 1$ as expected.
Indeed, in our toy model, the evolution outside the bounce phase is governed by a massless scalar field whose sound speed is the speed of light.
In general, the sound speed outside the NEC-violating phase will be that of the matter content dominating the evolution.
In particular, any canonical scalar field will give a sound speed equal to unity.
This avoids particular strong coupling issues that arise outside the bounce phase or for nearby background trajectories in phase space \cite{Dobre:2017pnt}.

\subsection{General evolution of the infrared perturbations in the bounce phase}

Now that we have assessed the stability of the model through the bounce phase, let us explore in more detail how the perturbations behave through the bounce, at least in the IR regime, i.e., for modes that are of observational interest in the expanding phase after the bounce.
In particular, we would like to address whether modes remain constant on super-Hubble scales or whether some amplification is experienced.
Earlier analyses (see, e.g., Refs.~\cite{Battarra:2014tga,Quintin:2015rta}) have indicated that IR modes tend to remain constant through a bounce phase, but some limited amplification might be allowed depending on the model parameters.
Nevertheless, if this amplification is too large, it could have important consequences such as the overproduction of non-Gaussianities \cite{Quintin:2015rta}.

The equation of motion for curvature perturbations \eqref{eq:zetaEOM} has the following super-Hubble limit\footnote{We omit the subscript $k$ since we work in the limit $k\rightarrow 0$, but it should be kept in mind that $\zeta$ is truly the Fourier transformed quantity $\zeta_k$. Also, we use the equality sign to lighten the notation, but all expressions in this subsection are actually given to leading order in the asymptotic limit $k/a\ll |H|$.}:
\begin{equation}
 \ddot\zeta+\left(\frac{\dot a}{a}+2\frac{\dot z}{z}\right)\dot\zeta=0\,,
\end{equation}
hence solutions are of the form\footnote{In this subsection, a subscript on a time-dependent quantity means that this quantity is evaluated at the corresponding time, e.g., $\zeta_0\equiv \zeta(t_0)$, $\dot\zeta_0\equiv\dot\zeta(t_0)$, $a_0\equiv a(t_0)$, $z_0\equiv z(t_0)$, etc.}
\begin{equation}
 \zeta(t)=\zeta_0+\dot\zeta_0\int_{t_0}^t\mathrm{d}\tilde t\,\left(\frac{a_0}{a(\tilde t)}\right)\left(\frac{z_0}{z(\tilde t)}\right)^2\,.
\end{equation}
In the context of a non-singular bounce occurring in the time interval $t_{B-}\leq t\leq t_{B+}$, the initial time would correspond to $t_{B-}$.
One can always rescale the time coordinate such that $t=t_B\equiv 0$ corresponds to the bounce point, and in a symmetric bounce, one can then set $t_{B+}=-t_{B-}$.
Consequently, one has $a_{B-}=a_{B+}$, $|H_{B-}|=|H_{B+}|$, and $\dot H_{B-}=\dot H_{B+}$.
The assumption that the bounce phase is symmetric in time can be relaxed, but it makes the analysis below considerably simpler for the sake of presentation.
We are interested in fluctuating modes that are generated deep in the contracting phase, that cross the Hubble radius still in the contracting phase and then enter the bounce phase as IR modes with amplitude $\zeta_{B-}$ and growth (or decaying\footnote{For a phase of contraction dominated by a canonical scalar field in Einstein gravity, the super-Hubble perturbations have a constant mode and either a growing or decaying mode depending on the equation of state of the field (see, e.g., Ref.~\cite{Quintin:2015rta}): for dust and radiation, the second mode grows; for a stiff fluid (e.g., a massless scalar), the second mode grows logarithmically in time; and for an ultra-stiff fluid (e.g., an Ekpyrotic field), the mode decays.}) rate $\dot\zeta_{B-}$.
The value of the curvature perturbation after the bounce phase, $\zeta_{B+}$, can then be expressed as
\begin{equation}
 \frac{\Delta\zeta}{\zeta_{B-}}=\frac{\dot\zeta_{B-}}{\zeta_{B-}}a_{B-}z^2_{B-}\int_{t_{B-}}^{t_{B+}}\frac{\mathrm{d}t}{a(t)z^2(t)}\,,
\end{equation}
where $\Delta\zeta\equiv\zeta_{B+}-\zeta_{B-}$. The question is then whether $\zeta$ can undergo an important growth through the bounce phase, i.e., whether $|\Delta\zeta/\zeta_{B-}|\gg 1$.

Before answering this question in the context of the model of this paper, let us make a comparison with a toy model: suppose the bounce phase was due to a field with constant equation of state, violating the NEC within GR.
In this crude model (which would presumably have ghost modes), one would find $z^2\propto a^2$ throughout the bounce regime.
Parameterizing the evolution in the bounce phase with $a(t)=a_B\mathrm{exp}(\Upsilon t^2/2)$ (so that $H(t)=\Upsilon t$), i.e., assuming this would be the solution to the equations of motion in such a bounce phase, one would find $\Delta\zeta\propto\int\mathrm{d}t\,\mathrm{exp}(-3\Upsilon t^2/2)$, the later being equal to the error function, an approximately linear function close to the origin.
Consequently, one could say that
\begin{equation}
 \left|\frac{\Delta\zeta}{\zeta_{B-}}\right|<\left|\frac{\dot\zeta_{B-}}{\zeta_{B-}}\right|\left(\frac{a_{B-}}{a_B}\right)^3\Delta t\,,
 \label{eq:comparconstEoS}
\end{equation}
where $\Delta t\equiv t_{B+}-t_{B-}=2|t_{B-}|$.
We note that $(a_{B-}/a_B)^3=1+3\Upsilon t_{B-}^2/2+...=1+\mathcal{O}(\Upsilon\Delta t^2)$,
so as long as the bounce phase is short compared to the scale $\dot H=\Upsilon$, the ratio $(a_{B-}/a_B)^3$ cannot be much larger than unity.
Accordingly, we see that, in this context, no significant amplification of the curvature perturbations through the bounce phase ($|\Delta\zeta/\zeta_{B-}|<1$) is ensured if $\Delta t^{-1}>|\dot\zeta_{B-}/\zeta_{B-}|$, i.e., if the time scale of the bounce phase is small compared to the fractional growth rate of the curvature perturbations when they enter the bounce phase.
Equation \eqref{eq:comparconstEoS} will serve as our basis for comparison with what happens in the model of this paper.

For cuscuton gravity, the expression for $z^2$ given in Eq.~\eqref{eq:z2gen} has the following IR limit:
\begin{equation}
 z^2=2a^2\frac{3\tilde Y}{3H^2+\dot H+\tilde Y}\,.
 \label{eq:z2IR}
\end{equation}
Recalling Eq.~\eqref{eq:bouncebckgrndseries}, we notice that $a_B\equiv a(t_B=0)=(\dot\chi_\mathrm{ini}/\sqrt{2|V_0|})^{1/3}$, hence
\[
 \tilde Y=\frac{\dot\chi^2}{2\mpl^2}=\frac{\dot\chi_\mathrm{ini}^2}{2\mpl^2a^6}=\frac{|V_0|}{\mpl^2}\left(\frac{a_B}{a}\right)^6\,.
\]
From this, Eq.~\eqref{eq:z2IR} can be rewritten as
\begin{equation}
 \frac{1}{az^2}=\frac{1}{6a^3}+\frac{\mpl^2}{6|V_0|}\frac{a^3}{a_B^6}(3H^2+\dot H)\,.
 \label{eq:1oaz2s}
\end{equation}
The first term above is reminiscent of the constant equation of state toy model with $z^2\propto a^2$, but the additional terms represent new contributions.
To evaluate the integral of the above quantity, and in particular to obtain an upper bound, we make use of certain properties of the functions $a(t)$, $H(t)$, and $\dot H(t)$ through the bounce.
Using the properties of the background solution \eqref{eq:bouncebckgrndseries} near the bounce point and extrapolating the zeroth-order expressions to the whole bounce phase, one can see that approximating the Hubble parameter to grow linearly as a function of time and $\dot H$ to be constant throughout the bounce phase amounts to finding an upper bound on the integral (i.e., overestimating it) as
\begin{equation}
 \int_{t_{B-}}^{t_{B+}}\frac{\mathrm{d}t}{a(t)z^{2}(t)}<\Delta t\max_{t\in[t_{B-},t_{B+}]}\frac{1}{a(t)z^2(t)}\,.
 \label{eq:intub}
\end{equation}
The maximum of $a(t)$, $a(t)^{-1}$, $H^2(t)$, and $\dot H(t)$ in the time interval $t\in[t_{B-},t_{B+}]$ is given by $a_{B-}$, $a_B$, $H_{B-}^2$, and $\dot H_B$, respectively.
Thus, applying Eq.~\eqref{eq:intub} with Eq.~\eqref{eq:1oaz2s} one finds
\begin{align}
 a_{B-}z^2_{B-}\int_{t_{B-}}^{t_{B+}}\frac{\mathrm{d}t}{a(t)z^2(t)}
 =&~\frac{6a_{B-}^3}{1+(\mpl^2/|V_0|)(a_{B-}/a_B)^6(3H_{B-}^2+\dot H_{B-})}
 \int_{t_{B-}}^{t_{B+}}\mathrm{d}t\,\left(\frac{1}{6a^3}+\frac{\mpl^2}{2|V_0|}\frac{a^3}{a_B^6}H^2+\frac{\mpl^2}{6|V_0|}\frac{a^3}{a_B^6}\dot H\right) \nonumber \\
 <&~\frac{6a_{B-}^3}{1+3(\mpl^2/|V_0|)(a_{B-}/a_B)^6H_{B-}^2}
 \left(\frac{1}{6a_B^3}+\frac{\mpl^2}{2|V_0|}\frac{a_{B-}^3}{a_B^6}H_{B-}^2+\frac{\mpl^2}{6|V_0|}\frac{a_{B-}^3}{a_B^6}\dot H_B\right)\Delta t\,,
\end{align}
where we make use of the fact that $\dot H_{B-}=0$ at the onset of the NEC-violating phase.
Using also the fact that $\dot H_B=|V_0|/|\mplt^2|$ (recall $V_0<0$ and $\mplt^2<0$) from Eq.~\eqref{eq:bouncebckgrndseries} at the bounce point, we obtain in the end
\begin{equation}
 \left|\frac{\Delta\zeta}{\zeta_{B-}}\right|<\left|\frac{\dot\zeta_{B-}}{\zeta_{B-}}\right|\left(\frac{a_{B-}}{a_B}\right)^3\Delta t\left[\frac{1+3\frac{\mpl^2}{|\mplt^2|}\left(\frac{a_{B-}}{a_B}\right)^3\left(\frac{H_{B-}^2}{\dot H_B}+\frac{1}{3}\right)}{1+3\frac{\mpl^2}{|\mplt^2|}\left(\frac{a_{B-}}{a_B}\right)^6\frac{H_{B-}^2}{\dot H_B}}\right]\,.
 \label{eq:Deltazetarescusc}
\end{equation}
In comparison with the toy model earlier, which gave the bound \eqref{eq:comparconstEoS} as a result, we see that the new term in the far right-hand side (the one in square brackets) is the one that determines whether the curvature perturbations can undergo significant amplification or not.
First, note that with the mass parameters in the range \eqref{eq:boundm2}, i.e., $1/2<m^2\mpl^2/M_L^4\leq 1$, one finds $1/2<\mpl^2/|\mplt^2|\leq 2$ from Eq.~\eqref{eq:defmplt}.
For our purpose here, this means $\mpl^2/|\mplt^2|=\mathcal{O}(1)$.
Therefore, one is left with three regimes to analyze.
If $H_{B-}^2/\dot H_B\gg 1$, then the term of Eq.~\eqref{eq:Deltazetarescusc} in square brackets becomes approximately $(a_B/a_{B-})^3$, which is smaller than unity.
Another regime corresponds to when $H_{B-}^2/\dot H_B\ll 1$, in which case the term of interest becomes approximately $1+(\mpl^2/|\mplt^2|)(a_{B-}/a_B)^3$, which can only be significantly larger than unity if the ratio $(a_{B-}/a_B)^3$ is.
However, this cannot be the case if $H_{B-}^2/\dot H_B\lesssim 1$.
To see this, reusing the approximation $H(t)=\Upsilon t+\mathcal{O}(\Upsilon^{3/2}t^2)$ as a proxy, one has $H_{B-}^2/\dot H_B=\Upsilon\Delta t^2/4+\mathcal{O}(\Upsilon^2\Delta t^4)\lesssim 1$, which justifies the expansion in $\Upsilon\Delta t^2\ll 1$.
Then, one immediately gets $(a_{B-}/a_B)^3=1+\mathcal{O}(\Upsilon\Delta t^2)$, which cannot be larger than $\mathcal{O}(1)$.
Finally, one has the regime where $H_{B-}^2/\dot H_B\sim 1$, but in that case, the term in square brackets above is necessarily $\mathcal{O}(1)$.
Consequently, across the range of possible values for the ratio $H_{B-}^2/\dot H_B$, the term in square brackets in Eq.~\eqref{eq:Deltazetarescusc} can never be larger than $\mathcal{O}(1)$.

To summarize, our results indicate that curvature perturbations in the IR do not undergo significantly more amplification through a bounce driven by a cuscuton field than through a toy-model bounce with constant phantom equation of state. Accordingly, provided $|\dot\zeta_{B-}/\zeta_{B-}|$ at the onset of the bounce phase is smaller than the inverse time duration of the bounce phase, curvature perturbations do not undergo significant amplification, and the constant mode solution on super-Hubble scales remains dominant throughout.
An immediate implication is that the production of non-Gaussianities through the bounce phase should not be catastrophic, contrary to what happens when $\zeta$ grows significantly \cite{Quintin:2015rta}, in particular significantly more compared to the constant equation of state toy model.
Yet, the proper study of non-Gaussianities in cuscuton gravity remains a topic of future work.

\section{The case of extended cuscuton}\label{sec:extendedcuscuton}

\subsection{Introducing a concrete extended cuscuton model}

Earlier, we introduced the cuscuton as a scalar field that satisfies at most a first-order constraint equation at the background level and that has no kinetic term at the perturbation level.
This was implemented within $k$-essence theory, but this can be generalized.
As done in Ref.~\cite{Iyonaga:2018vnu}, one can instead start from the more general Horndeski theory (or even beyond-Horndeski theory) and similarly derive conditions on the free functions so that the scalar field does not propagate both at the background and perturbation level.
This can also be approached from a Hamiltonian construction \cite{Mukohyama:2019unx}, demanding the theory to propagate only two degrees of freedom (the two tensor modes of GR).
Out of this arises the class of `extended' cuscuton.

While the extended cuscuton represents a large class of models, in this subsection we focus on a specific example and explore its consequences.
Specifically, let us consider the following simple subclass of extended cuscuton,
\begin{align}
 S = \int \mathrm{d}^4 x \sqrt{-g}\,\left[
\frac{\mpl^2}{2} R - M_L^2\sqrt{2X} -V(\phi) - \lambda \left(\frac{3 \lambda }{\mpl^2} (2X) - \ln\left(\frac{2X}{\Lambda^4}\right) \Box \phi \right)  - \frac{1}{2} g^{\mu\nu}\partial_{\mu}\chi \partial_{\nu} \chi
\right],
\end{align}
where $\Box\phi\equiv g^{\mu\nu}\nabla_\mu\nabla_\nu\phi$.
In the above, $\lambda$ and $\Lambda$ are new parameters with dimensions of mass. The original cuscuton model corresponds to\footnote{This action corresponds to the following parameters of Ref.~\cite{Iyonaga:2018vnu} up to a total derivative:
$\tilde{b}_0(\phi) = \tilde{v}_4(\phi) = \mpl^2/2$, 
$\tilde{u}_2(\phi) = - V(\phi)$,
$\tilde{v}_2(\phi) = -M_L^2$,
$\tilde{v}_3(\phi) = - 2 \lambda$,
$\tilde{b}_1(\phi) = \tilde{u}_4(\phi) = 0$,
and $A_5(\phi,X) = 0$.} $\lambda = 0$.
In an FLRW background, the equations of motion reduce to the following set of independent ordinary differential equations:
\begin{align}
 3\mpl^2\Theta^2&=\frac{1}{2}\dot\chi^2+V(\phi)\,,\\
 3M_L^2\Theta&=V_{,\phi}-\frac{6\lambda}{\mpl^2}V(\phi)\,, \label{eq:constraintextended}\\
 \ddot\chi+3H\dot\chi&=0\,,
\end{align}
where $\Theta$ is defined by\footnote{Note that our definition of $\Theta$ differs from that of Ref.~\cite{Iyonaga:2018vnu} by a factor of $\mpl^2$. Rather than working with $\Theta$ having dimension of mass cubed, it has dimension of mass here. It is easier to think of $\Theta$ as a shifted Hubble parameter that way.}
\begin{equation}
 \Theta\equiv H+\frac{\lambda}{\mpl^2}\dot\phi\,.
\end{equation}
For completeness, these equations can be combined to yield the second Friedmann equation $2\mpl^2\dot H=-\dot\chi^2+M_L^2\dot\phi-2\lambda(\ddot\phi-3\Theta\dot\phi)$, or in terms of $\Theta$,
\begin{equation}
 2\mpl^2\dot\Theta=-\dot\chi^2+(M_L^2+6\lambda\Theta)\dot\phi\,.
 \label{eq:2ndFriedmannextended}
\end{equation}
We note that, here again, we focus on the case $\dot{\phi} > 0$.
Also, note that $\Lambda$ does not affect the dynamics because it is just a total derivative in the action.
One notices that the above equations are very similar to the background equations for the standard cuscuton model [Eqs.~\eqref{Friedconstraint}--\eqref{eq:cusconstreqFLRW2}] with the Hubble parameter replaced by a `shifted' Hubble parameter $\Theta$.
Indeed, it is this quantity that is bounded, provided $V_{,\phi}-6\lambda V(\phi)/\mpl^2$ remains finite for all values of $\phi$.
Hence the extended cuscuton model presented here is still a limiting curvature theory, allowing for non-singular solutions.
The particularity here is that there can be a point in time where $\Theta=0$, more specifically if $H=-\lambda\dot\phi/\mpl^2$ in a contracting phase.
As we will see below, the point where $\Theta=0$ becomes a `critical' crossing time, which we call `$\Theta$ crossing' in a similar fashion to Hubble crossing, where one has to be careful depending on the gauge used.
It also corresponds to the so-called $\gamma$ (or\footnote{The variables $\gamma$ and $\Theta$ are used interchangeably in the literature (see, e.g., Refs.~\cite{Kobayashi:2016xpl,gammacrossing}).} $\Theta$) crossing encountered in general non-singular cosmology within Horndeski theory (and beyond).
We do not construct a full solution here with $\Theta$ crossing, but we note that it is straightforward to see that there exist such solutions.
Indeed, a similar background solution to what was found in Sec.~\ref{sec:cuscutonbounce} exists, except $H$ crossing is replaced by $\Theta$ crossing, i.e., $\Theta=0$ when $\dot\chi^2/2=-V(\phi)$, hence $V(\phi)$ is negative at $\Theta$ crossing.
As one can see from Eq.~\eqref{eq:constraintextended}, a further requirement on the potential here is that $V_{,\phi}=6\lambda V(\phi)/\mpl^2$ at $\Theta$-crossing time.
A non-singular bounce would follow at a later time, with Hubble crossing happening when $\Theta=\lambda\dot\phi/\mpl^2$.

\subsection{Finiteness and stability of the cosmological perturbations}

Let us explore the cosmological perturbations of the extended cuscuton model presented in the last subsection and comment on the (in)validity of certain gauges at $\Theta$ crossing and the general stability of the theory.
Following the methodology of the previous section, let us evaluate the second-order action written in terms of gauge-invariant variables based on the spatially-flat gauge.
Similarly to the minimal cuscuton case, $\Phi^{S}$ and $B^{S}$ do not have kinetic terms, and from the variations of these fields we obtain\footnote{The action is expanded in Fourier space, hence all perturbation variables written below represent their Fourier transform. We simply omit the subscript $k$ to lighten the notation in this section.}
\begin{align}
\Phi^S &= \frac{1}{2 \mpl^2 \Theta} \left( \dot{\chi} \delta\chi^S - (M_L^2+6\lambda\Theta) \delta \phi^S + 2 \lambda \dot{\delta\phi}^S \right)\,,\label{constext1} \\
a B^S &= -\frac{\lambda}{\mpl^2 \Theta} \delta \phi^S + \frac{a^2}{2 k^2 \mpl \Theta^2} \left[\sqrt{2\tilde Y}\left(
(3\Theta^2-\tilde Y)\delta\chi^S+\Theta\dot{\delta\chi}^S\right)
+\frac{\tilde Y}{\mpl}\left(M_L^2\delta\phi^S-2\lambda\dot{\delta\phi}^S\right)
\right]\,.\label{constext2}
\end{align}
After plugging the above expressions into the second-order action, it becomes a function of two variables: $\delta \chi^S$ and $\delta \phi^S$.
Though both fields appear to have kinetic terms, they are degenerate.
The kinetic term can be diagonalized by introducing $\zeta$ instead of $\delta \chi^S$ as follows:
\begin{align}
 \zeta\equiv-\frac{\Theta}{\dot\chi}\delta\chi^S+\frac{\lambda}{\mpl^2}\delta\phi^S\,.
 \label{defzeta}
\end{align}
Then the cuscuton field has no kinetic term, and one obtains the following constraint equation:
\begin{equation}
 \delta\phi^S=\frac{2M_L^2\mpl^2\tilde Y(\tilde Y+\dot{\Theta})\left[\left(M_L^2(\tilde Y+\dot\Theta)+3\Theta^2(M_L^2+6\lambda\Theta)\right)\zeta-\Theta\left(M_L^2+6\lambda\Theta\right)\dot\zeta\right]}{\left[M_L^4\tilde Y(\tilde Y+\dot\Theta)+\Theta^2(M_L^2+6\lambda\Theta)\left((M_L^2+8\lambda\Theta)k^2/a^2+3M_L^2\tilde Y\right)\right]\left(2\lambda(\tilde Y+\dot{\Theta})-\Theta(M_L^2+6\lambda\Theta)\right)}\,.
\end{equation}
Upon eliminating $\delta\phi^{S}$ from the perturbed action, one finds
\begin{align}
 S=\frac{\mpl^2}{2}\int\mathrm{d}t\mathrm{d}^3\mathbf{k}~az^2\left(\dot\zeta^2-c_S^2 \frac{k^2}{a^2}\zeta^2\right)\,.
\end{align}
The functions $z^2$ and $c_S^2$ are given by
\begin{align}
 z^2&=2a^2\tilde Y\left(\Theta^2+\frac{M_L^4\tilde Y(\tilde Y+\dot{\Theta})}{(M_L^2+6\lambda\Theta)\left((M_L^2+8\lambda\Theta)k^2/a^2+3M_L^2\tilde Y\right)}\right)^{-1}\,, \nonumber \\
 c_S^2&=\frac{A_4(k/a)^4+A_2(k/a)^2+A_0}{B_4(k/a)^4+B_2(k/a)^2+B_0}\,,
\end{align}
where we define
\begin{equation}
 A_4\equiv B_4\equiv\Theta^2(M_L^2+6\lambda\Theta)^3(M_L^2+8\lambda\Theta)^2\,,
\end{equation}
and
\begin{align}
 A_2\equiv&~M_L^2(M_L^2+6\lambda\Theta)(M_L^2+8\lambda\Theta)
 \Biggl[\tilde Y\left(12\Theta^2(M_L^2+6\lambda\Theta)(M_L^2+7\lambda\Theta)+M_L^2(3M_L^2+20\lambda\Theta)\dot\Theta\right) \nonumber \\
    &+\tilde Y^2\left(M_L^4-2\lambda\Theta(7M_L^2+72\lambda\Theta)\right)
    +(M_L^2+8\lambda\Theta)\left(-\Theta(M_L^2+6\lambda\Theta)\ddot\Theta+2(M_L^2+9\lambda\Theta)\dot\Theta^2\right)\Biggr]\,, \nonumber \\
 A_0\equiv&~M_L^4\tilde Y\Biggl[16M_L^2\lambda^2\tilde Y^3-\tilde Y^2(M_L^2+6\lambda\Theta)(M_L^4+50M_L^2\lambda\Theta+240\lambda^2\Theta^2)+3\tilde Y\Theta^2(M_L^2+6\lambda\Theta)^2(5M_L^2+34\lambda\Theta) \nonumber \\
    &+\dot{\Theta}\left(48M_L^2\lambda^2\tilde Y^2+\tilde Y(M_L^2+6\lambda\Theta)(M_L^4+20M_L^2\lambda\Theta+240\lambda^2\Theta^2)-12\Theta^2(M_L^2+6\lambda\Theta)^2(M_L^2+8\lambda\Theta)\right) \nonumber \\
    &+2\dot\Theta^2\left(M_L^6+24M_L^2\lambda^2\tilde Y+\lambda\Theta(41M_L^4+450M_L^2\lambda\Theta+1440\lambda^2\Theta^2)\right)+16M_L^2\lambda^2\dot\Theta^3 \nonumber \\
    &-3\Theta(M_L^2+6\lambda\Theta)^2(M_L^2+8\lambda\Theta)\ddot\Theta\Biggr]\,, \nonumber \\
 B_2\equiv&~M_L^2\tilde Y(M_L^2+8\lambda\Theta)(M_L^2+6\lambda\Theta)^2\left[M_L^2(\tilde Y+\dot\Theta+6\Theta^2)+36\lambda\Theta^3\right]\,, \nonumber \\
 B_0\equiv&~3M_L^4\tilde Y^2(M_L^2+6\lambda\Theta)^2\left[M_L^2(\tilde Y+\dot\Theta+3\Theta^2)+18\lambda\Theta^3\right]\,.
\end{align}
One notices that, as long as $\Theta\neq 0$, the kinetic and gradient terms have the correct sign for large-$k$ modes:
\begin{align}
 z^2&=\frac{2a^2\tilde Y}{\Theta^2}+\mathcal{O}\left(\frac{a^2}{k^2}\right)>0\,; \nonumber \\
 c_S^2&=1+\mathcal{O}\left(\frac{a^2}{k^2}\right)>0\,.
\end{align}
In fact, $z^2>0$ for all $k$ modes as long as one imposes $M_L^2+8\lambda\Theta>0$.
Indeed, if that is the case, $M_L^2+6\lambda\Theta>0$ is immediately ensured, and so is $\tilde Y+\dot\Theta=(M_L^2+6\lambda\Theta)\dot\phi/2\mpl^2>0$ according\footnote{Recall also that $\dot\phi>0$ by assumption in this section.} to Eq.~\eqref{eq:2ndFriedmannextended}.
This requirement, rewritten as $\Theta>-M_L^2/8\lambda$, sets the minimal value (or maximally negative value) $\Theta$ can take.
Since $\dot\phi>0$, one has $\Theta>H$, hence $\mathrm{min}(\Theta)>\mathrm{min}(H)=-\mathrm{max}(|H|)=-|H_{B-}|$.
Therefore, one needs the curvature scale of the bounce, of the order of $|H_{B-}|$, to be at most of the order of $M_L^2/\lambda$, essentially the new limiting curvature scale, which is a reasonable assumption.

Similarly to the issue of Hubble crossing in the previous section, one might worry about what happens when $\Theta=0$ here.
Before looking at any particular wavenumber limit, the important point is that the coefficients must remain well defined at $\Theta$ crossing.
We can see that this is the case by expanding about $\Theta=0$:
\begin{align}
 z^2&=2a^2\frac{k^2/a^2+3\tilde Y}{\tilde Y+\dot\Theta}+\mathcal{O}(\Theta)\,; \nonumber \\
 c_S^2&=\frac{M_L^4\left(\tilde Y+2\dot\Theta\right)k^2/a^2+\tilde Y\left(\tilde Y\left(-M_L^4+16\lambda^2\tilde Y\right)+2\dot\Theta\left(M_L^4+16\lambda^2\tilde Y\right)+16\lambda^2\dot\Theta^2\right)}{M_L^4\tilde Y(k^2/a^2+3\tilde Y)}+\mathcal{O}(\Theta)\,.
\end{align}
Note that $z^2$ is always positive because, according to the background equation \eqref{eq:2ndFriedmannextended},
\[
 \tilde Y+\dot\Theta=\frac{M_L^2}{2\mpl^2}\dot\phi\,,\qquad\mathrm{when}\ \Theta=0\,,
\]
and so $\tilde Y+\dot\Theta$ is positive following our assumption that $\dot\phi>0$.
For large (but finite) $k$ modes, the sound speed at $\Theta$ crossing is found to be
\begin{equation}
 c_S^2=1+\frac{2\dot\Theta}{\tilde Y}+\mathcal{O}(k^{-2}, \Theta)\,,
\end{equation}
and hence gradient instabilities and superluminality can be avoided if $-\tilde  Y/2<\dot\Theta\leq 0$, or in terms of the original variables,
\begin{equation}
 -\frac{\dot\chi^2}{4\mpl^2}<\dot H+\frac{\lambda}{\mpl^2}\ddot\phi\leq 0\,,
\end{equation}
when $\Theta=0$.
When this is satisfied, one can conclude that the curvature perturbation $\zeta$ is well behaved at $\Theta$ crossing, at least in the UV.
We do not check the above condition explicitly for a given background solution, but there is \textit{a priori} no reason why this could not be satisfied.

Finally, we want to verify whether or not the metric and scalar field perturbations remain finite at $\Theta$ crossing.
At first glance, it looks like diverges cannot be avoided because of $1/\Theta$ and $1/\Theta^2$ factors appearing in the constraint equations \eqref{constext1} and \eqref{constext2}.
However, one can show that all divergences actually cancel out, and one rather obtains finite results as follows:
\begin{align}
 \Phi^S=&~\frac{1}{4M_L^4\lambda^2(\tilde Y+\dot\Theta)^2(k^2/a^2+3\tilde Y)}\Biggl\{3M_L^4\tilde Y\left[-24\lambda^2\tilde Y^2+\left(M_L^4-12\lambda^2\tilde Y\right)\dot\Theta+12\lambda^2\dot\Theta^2-2M_L^2\lambda\ddot\Theta\right] \nonumber \\
    &+\frac{k^2}{a^2}\biggl[8\lambda^2\tilde Y^2\left(-5M_L^4+8\lambda^2\tilde Y\right)+\left(M_L^8-44M_L^4\lambda^2\tilde Y+192\lambda^4\tilde Y^2\right)\dot\Theta \nonumber \\
    &+4\lambda^2\left(-M_L^4+48\lambda^2\tilde Y\right)\dot\Theta^2+64\lambda^4\dot\Theta^3-2M_L^6\lambda\ddot\Theta\biggr]\Biggr\}\zeta \nonumber \\
    &+\frac{\tilde Y}{2M_L^2\lambda(\tilde Y+\dot\Theta)^2}\left(M_L^4-\frac{16\lambda^2(k/a)^2\left(\tilde Y+\dot\Theta\right)}{(k/a)^2+3\tilde Y}\right)\dot{\zeta}+\mathcal{O}(\Theta)\,, \nonumber \\
 aB^S=&-\frac{(k/a)^2[M_L^4-16\lambda^2(\tilde Y+\dot\Theta)]\zeta+6M_L^2\lambda\tilde Y\dot\zeta}{2M_L^2\lambda(k/a)^2(\tilde Y+\dot\Theta)}+\mathcal{O}(\Theta)\,, \nonumber \\
 \delta\phi^S=&~\frac{\mpl^2}{\lambda}\zeta+\mathcal{O}(\Theta)\,, \nonumber \\
 \delta\chi^S=&\frac{\mpl\sqrt{2\tilde Y}}{2\lambda(\tilde Y+\dot{\Theta})}(M_L^2\zeta-2\lambda\dot\zeta)+\mathcal{O}(\Theta)\,.
\end{align}
Therefore, while perturbations appear pathological at first sight when $\Theta=0$, it turns out not to be the case. Rather, all cosmological perturbations are well defined here in the spatially-flat gauge, and the results can be trusted, i.e., the fact that there can be no ghost and no gradient instability is robust.
This conclusion extends to the other common gauges explored in this work (Newtonian gauge, comoving gauge with respect to the cuscuton field and with respect to the matter field).
We do not write down the expression for all the perturbation variables in the different gauges here, but one can see from Eqs.~\eqref{eq:Stophigauges}, \eqref{eq:Stochigauges} and \eqref{eq:Ngtransfs} that the transformations from the spatially-flat gauge to the other gauges do not introduce any divergence when $\Theta=0$.
There might remain other points in time where certain gauges are not well defined, but as before, pathologies are often resolved when approached carefully or when transforming to another gauge, so they do not represent physical issues.

\section{Discussion and conclusions}\label{sec:fin}

Let us summarize the results of this work.
We began by reviewing cuscuton gravity as a minimal modification to GR, where one introduces a scalar field that does not propagate any new degrees of freedom on a cosmological background.
Correspondingly, the scalar field satisfies a first-order constraint equation at the background level and the perturbed action in the scalar sector has no kinetic term when no additional matter is included.
It was shown how cuscuton gravity naturally implements the concept of limiting curvature by bounding the trace of the extrinsic curvature.
In particular, using cuscuton gravity with the addition of a massless scalar field, the background solution of a non-singular bouncing cosmology was explicitly obtained.

The cosmological perturbations were then revisited.
We confirmed that the model is free of ghost instabilities since the sign of the kinetic term is always positive.
The sign of the sound speed squared is also positive, in particular in the UV limit.
Interestingly at the bounce point, i.e., when the Hubble parameter crosses zero, the sound speed becomes greater than the speed of light.
In the IR, requiring the sound speed square to be positive constrains the model parameters (the limiting curvature scale of the cuscuton field $M_L$ and the value of the second derivative of the cuscuton potential at the bounce point, denoted $m^2$) to lie in some reasonable range.
In particular, a sound speed close to unity is ensured when $m^2\approx M_L^4/\mpl^2$.
This also sets $c_S^2$ close to the bounce point for modes with $k/a\gtrsim\mathcal{O}(\dot\chi/\mpl)$ to be at most $\mathcal{O}(10)$.
The evolution of IR perturbations was then studied in greater detail.
We found that curvature perturbations can receive only limited growth through the bounce phase.
If the fractional growth rate of the curvature perturbations when they enter the bounce phase is small compared to the inverse time duration of the bounce phase, then the constant mode solution on super-Hubble scales remains the dominant solution throughout the bounce phase.

Special emphasis was given to exploring the perturbations in different gauges to confirm the validity of the perturbation equations and the aforementioned stability results.
We found that the spatially-flat gauge is ill defined when $H=0$, but the Newtonian gauge, the comoving gauge with respect to the cuscuton field and the comoving gauge with respect to the matter field are all well defined at that point.
Indeed, it was found that the divergence that could occur in the cosmological perturbations when $H=0$ exactly cancel out in these gauges.
Therefore, the stability results listed above were shown to be robust.
The Newtonian gauge has another potentially problematic crossing point when $k^2/a^2=3\dot H$ in a bounce phase.
However, apparent divergences at that point were shown to cancel out once again.

The analysis was brought one step further by considering a model of extended cuscuton gravity, which is a subclass of Horndeski theory.
In particular, we considered the addition of a cubic Galileon-like term proportional to $\Box\phi$, such that the vacuum theory still propagates only two degrees of freedom in the unitary gauge.
Including a massless scalar field once again, we derived the scalar cosmological perturbations in the spatially-flat gauge.
Stability was verified: the kinetic term has the correct sign and the sound speed is equal to unity to leading order in the UV.
While apparent divergences could arise when the Hubble parameter crossed zero in the original cuscuton model, in the extended cuscuton model the potentially problematic crossing time corresponds to $\Theta=0$, where $\Theta$ is a `shifted' Hubble parameter, a common parameter appearing in scalar-tensor theories with kinetic braiding.
In this case, we showed that all perturbations remain finite when $\Theta=0$, even in the spatially-flat gauge, therefore confirming the robustness of the stability results in a non-singular bouncing cosmology implementing limiting curvature.

This work opens up many interesting future directions.
A first notable topic that was not developed is related to the strong coupling issue.
Typically in a non-singular cosmology, strong coupling appears to be inevitable within $k$-essence theory \cite{deRham:2017aoj} or at best borderline avoidable within Horndeski theory \cite{Koehn:2015vvy,Dobre:2017pnt}.
First indications show that the sound speed in this work can remain far from the regimes $c_S^2\ll 1$ and\footnote{We note, however, that a superluminal sound speed with $c_S^2\gg 1$ can sometimes alleviate the strong coupling problem (see, e.g., Ref.~\cite{Joyce:2011kh}).} $c_S^2\gg 1$ throughout time.
Thus, the strong coupling issue that usually plagues NEC-violating $k$-essence theories might be evaded in cuscuton gravity.
Yet, this remains to be properly shown.
This is also related to the study of non-Gaussianities, which would be interesting to develop in the context of cuscuton gravity.

A curious result of this work is that the sound speed in cuscuton gravity can become slightly larger than the speed of light close to a bounce point.
Usually, theories admitting superluminal propagation cannot be UV-completed in a `standard' manner, i.e., they cannot possess an analytic S-matrix \cite{Adams:2006sv}.
This might not be surprising since the cuscuton Lagrangian involves the square root of $|X|$, which is a non-analytic function in the neighborhood of $X=0$, hence rendering quantization particularly challenging due to branch cuts at that point.
Another reason why this might not be too surprising comes from the fact that general theories of minimally-modified gravity often have to explicitly break Lorentz invariance (see, e.g., Ref.~\cite{Mukohyama:2019unx}), for instance breaking time diffeomorphism invariance but keeping 3-dimensional space-like diffeomorphisms.
Yet, other approaches to UV completion could be explored in this context (e.g., see \cite{classUV}), and this topic certainly deserves further attention.

Back to classical aspects of this work, we note that it would be interesting to explore how the implementation of limiting curvature in cuscuton gravity could be extended to more general theories of modified gravity.
Indeed, the fact that it is the trace of the extrinsic curvature of a 3-dimensional spatial hypersurface that is bounded in the theory, rather than a 4-dimensional space-time curvature-invariant quantity as in Refs.~\cite{Mukhanov:1991zn,Brandenberger:1993ef,Yoshida:2017swb,Yoshida:2018kwy}, is an indication that more limiting curvature theories could be constructed, in this case explicitly breaking time diffeomorphism invariance.
Moreover, this might pave the way to a better classification of limiting curvature theories, such as cuscuton gravity and mimetic gravity, and clarifying their relation.

\begin{acknowledgments}
We would like to thank Niayesh Afshordi, Jean-Luc Lehners and Jo\~ao Magueijo for stimulating discussions.
We also thank Robert Brandenberger for involvement in the initial stages of this project and for valuable comments.
Research at the Albert Einstein Institute (AEI) is supported by the European Research Council (ERC) in the form of the ERC Consolidator Grant CoG 772295 ``Qosmology''.
J.\,Q.~further acknowledges financial support (throughout the completion of this work) in part from
the Vanier Canada Graduate Scholarship administered by
the Natural Sciences and Engineering Research Council of Canada (NSERC),
the NSERC Postdoctoral Fellowship and
the \textit{Fond de recherche du Qu\'ebec \textendash\ Nature et technologies} postdoctoral research scholarship.
D.\,Y.~is supported by the Japan Society for the Promotion of Science (JSPS) Postdoctoral Fellowship No.~201900294 and thanks the AEI for hospitality during the completion of this work.
We also acknowledge the stimulating atmosphere at McGill University where this project started.
\end{acknowledgments}

\end{document}